\documentclass[%
superscriptaddress,
preprint,
 amsmath,amssymb,
 aps,
]{revtex4-2}

\usepackage{times}

\usepackage{amsmath}
\usepackage{amsfonts}
\usepackage{amssymb}
\usepackage{graphicx}
\usepackage{xcolor}

\usepackage[shortlabels]{enumitem}
\usepackage{multirow}

\usepackage[labelfont=bf]{caption}
\usepackage{setspace}
\newcommand{\beginsupplement}{%
        \setcounter{table}{0}
        \renewcommand{\thetable}{S\arabic{table}}%
        \setcounter{figure}{0}
        \renewcommand{\thefigure}{S\arabic{figure}}%
     }

\begin{document} 
 
\title{Universal self-similarity of hierarchical communities formed through a general self-organizing principle} 

\author{Shruti Tandon and Nidhi Dilip Sonwane}
\altaffiliation{These authors contributed equally}
\email{Corresponding author: shrutitandon97@gmail.com}
 \affiliation{%
 \small{Department of Aerospace Engineering, Indian Institute of Technology Madras, Chennai 600036, India}
} 
\affiliation{%
 \small{Centre of Excellence for Studying Critical Transitions in Complex Systems, Indian Institute of Technology Madras, Chennai 600036, India}
}

\author{Tobias Braun}

\affiliation{%
 \small{Centre of Excellence for Studying Critical Transitions in Complex Systems, Indian Institute of Technology Madras, Chennai 600036, India}
}
\affiliation{%
 \small{Institute for Earth System Science \& Remote Sensing, Leipzig University, Leipzig 04103, Germany}
} 

\author{Norbert Marwan}
\affiliation{%
 \small{Potsdam Institute for Climate Impact Research (PIK), Member of the Leibniz Association, 14412 Potsdam}
}
\affiliation{%
 \small{Institute of Geosciences, University of Potsdam, Potsdam 14476, Germany}
}

\author{J\"urgen Kurths}
\affiliation{%
 \small{Potsdam Institute for Climate Impact Research (PIK), Member of the Leibniz Association, 14412 Potsdam}
}
\affiliation{%
 \small{Institute of Physics, Humboldt Universit\"at zu Berlin, Berlin 12489, Germany}
}

\author{R. I. Sujith}
\affiliation{%
\small{ Department of Aerospace Engineering, Indian Institute of Technology Madras, Chennai 600036, India}
} 
\affiliation{%
 \small{Centre of Excellence for Studying Critical Transitions in Complex Systems, Indian Institute of Technology Madras, Chennai 600036, India}
}





\begin{abstract}
    Emergence of self-similarity in hierarchical community structures is ubiquitous in complex systems. Yet, there is a dearth of universal quantification and general principles describing the formation of such structures. Here, we discover universality in scaling laws describing self-similar hierarchical community structure in multiple real-world networks including biological, infrastructural, and social networks. We replicate these scaling relations using a phenomenological model, where nodes with higher similarity in their properties have greater probability of forming a connection. A large difference in their properties forces two nodes into different communities. Smaller communities are formed owing to further differences in node properties within a larger community. We discover that the general self-organizing principle is in agreement with Haken’s principle; nodes self-organize into groups such that the diversity (differences) between properties of nodes in the same community is minimized at each scale and the organizational entropy decreases with increasing complexity of the organized structure.
\end{abstract}

\maketitle 

\newpage

\section*{Introduction}

Real-world systems are often complex with intricate structural properties that emerge from the holistic effects of a network of numerous local interactions. Such systems in biology \cite{white1986structure,eisenberg2000protein,vishveshwara2002protein,brinda2005network,sistla2005identification,shoemaker2007deciphering,bullmore2009complex,sporns2011human,amunts2013bigbrain,safari2014protein}, ecology \cite{dawah1995structure,krause2003compartments,hill2008network,fernanda2019network} and social sciences \cite{newman2003social,newman2001scientific,newman2004coauthorship} have been extensively studied using network representation. The individual components of a complex system are represented by nodes of a network and their interactions are encoded as links between these nodes. This approach allows viewing diverse systems through a general mathematical framework and thus enables the study of ubiquitous features across systems. For example, diverse real-world networks exhibit similar scale-free patterns of connectivity \cite{barabasi1999emergence}.

Many other structural features observed in networks, such as grouping of nodes into communities \cite{newman2006modularity,newman2012communities}, hierarchical topology \cite{schaub2023hierarchical, simon1962architecture}, fractal patterns of connectivity \cite{song2005self,song2006origins,ikeda2021stratified,gallos2007review, dorogovtsev2002pseudofractal}, and phase transitions in growing networks \cite{strogatz2001exploring,boccaletti2006complex, dorogovtsev2008critical}, reveal the degree of complexity in the organization of complex systems. The fact that diverse systems can exhibit similar structural features incites the idea of universality; i.e., the emergence of such features is independent of the finer details of the system \cite{kwapien2012physical, siegenfeld2020introduction}. Then, universality in a quantitative description of such features, such as scaling laws, is as inevitable as intriguing. Any universal scaling law must indeed lead to general self-organizing principles ubiquitous across various complex systems. Here, we report the striking universality in scaling laws describing the self-similar topology of communities-within-communities formed in multiple real-world networks including social interaction, infrastructural, and biological networks. Further, we use a basic phenomenological model to explain the emergence of hierarchical communities obeying such universal scaling laws and discover the underlying self-organizing principle.

Many real-world networks exhibit a topology made up of communities. A community is formed when a group of nodes interact more among themselves than with nodes from any other group. Hierarchical communities can be formed when nodes in larger communities further sub-group into smaller communities at multiple scales \cite{simon1962architecture,schaub2023hierarchical}. Such hierarchical community structure is observed in many physical networks such as human brain networks \cite{meunier2009hierarchical, meunier2010modular}, infrastructural networks \cite{rosvall2011multilevel}, social networks \cite{newman2001scientific,barabasi2002evolution,ravasz2003hierarchical,newman2004coauthorship,flake2002self,onnela2007structure,guimera2003self}, biological networks \cite{hartwell1999molecular,rives2003modular,luo2007modular,ravasz2002hierarchical,lewis2010function,bullmore2009complex,zhao2023intrinsic,sporns2016modular,meunier2010modular} etc. 

\begin{figure}
    \centering
    \includegraphics[width=0.8\textwidth]{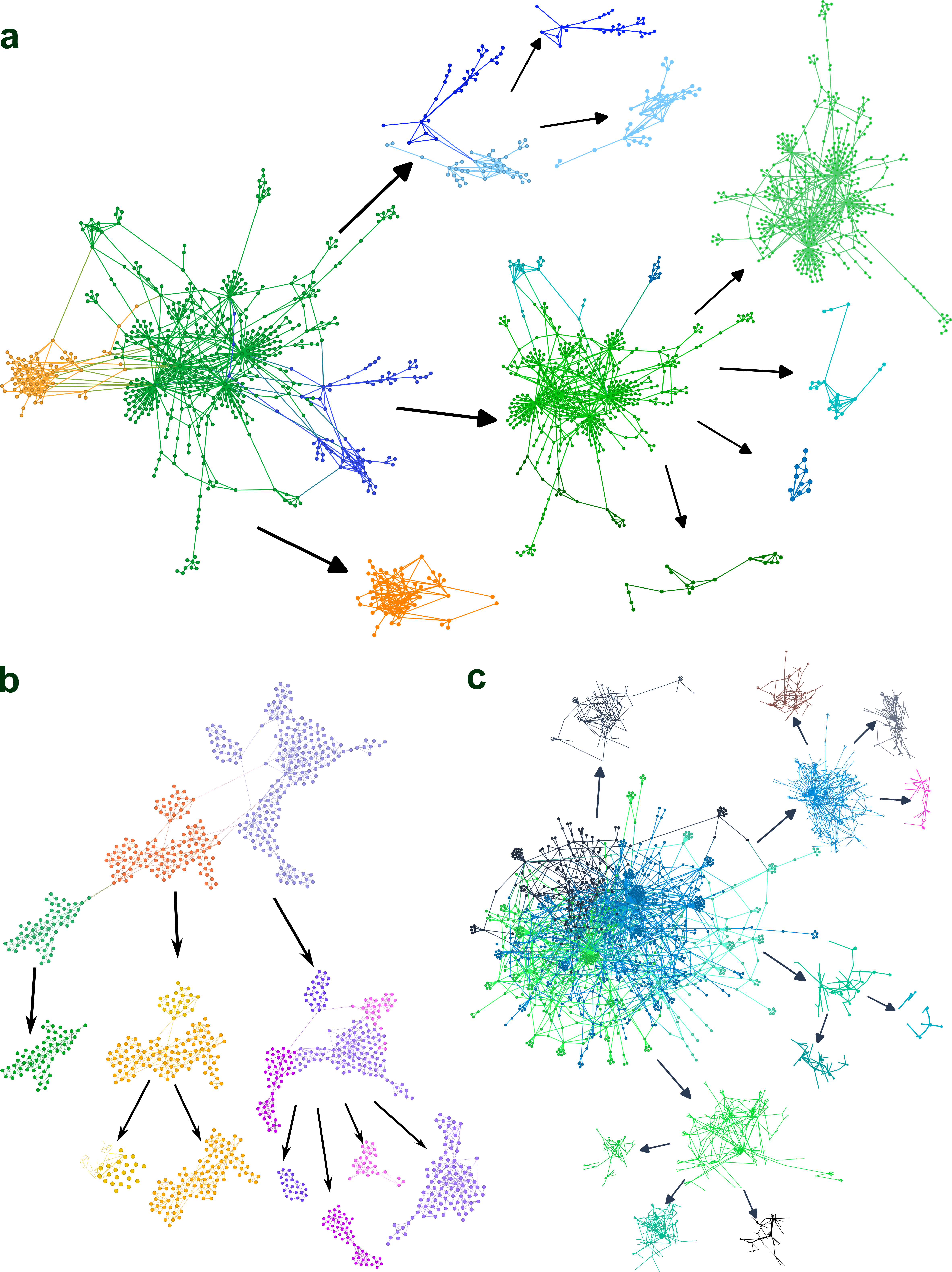}
    \caption{\textbf{Examples of real-world networks that self-organize into a hierarchical community structure.} An illustration of the community-within-community structure in \textbf{(a)} protein-protein interaction network of the bacterium \textit{Escherichia coli},  \textbf{(b)} network of coauthors working in network science and \textbf{(c)} network of nerve fibre tracts in mouse (visualized using Gephi \cite{bastian2009gephi} from openly available datasets \cite{chakrabarty2016naps, newman2006finding,bigbrain,nr}).}
    \label{fig_realntwk}
\end{figure}

Albeit originating from distinct physical systems, there is similarity in the multi-scale community structure in a network of protein-protein interaction in the bacterium \textit{Escherichia coli} \cite{peregrin2009modular, lewis2010function}, a co-authorship network of scientists working on network science and a network of nerve fibre tracts in mouse (see Fig. \ref{fig_realntwk}(a), (b) and (c), respectively). Hierarchical communities can also be topologically self-similar as known for networks of email-based social interactions \cite{guimera2003self}, scientific collaborations \cite{arenas2004community} and mammalian societies \cite{hill2008network}. That is, the topology of sub-grouping of a community into constituent communities is similar across multiple scales.

Emergence of complexity in the structural organization of complex systems is ubiquitous and yet perplexing. Self-similarity in hierarchical community structure of real-world networks is one such complexity that remains unexplained. The organization of a system into communities and hierarchies is beneficial for various reasons. The formation of communities helps a system to categorize sub-units according to their function and control the effect of associated errors \cite{fortunato2010community}. The impact of disturbances (such as the sudden removal of some nodes) can become contained within a community making the network robust against attacks or local failures \cite{krause2003compartments, fernanda2019network}. Moreover, connections between nodes in different communities govern the communication across communities and pave the way for decentralized functioning of the system \cite{strogatz2001exploring,krause2003compartments,siegenfeld2020introduction,fortunato2010community}. How does a complex system self-organize into such an optimal structure, and why do individual nodes comply with such an organization process remain important open questions.

Evidently, the structure of interactions between constituents is closely related to the inherent properties of each constituent \cite{strogatz2001exploring,siegenfeld2020introduction,ladyman2013complex,ottino2003complex,witherington2011taking}. For example, the structural connectivity in brain networks imposes restrictions on the functionalities of different components of the brain \cite{bullmore2009complex}. Nodes which are similar in terms of function/interest/origin tend to group together to form communities, and this understanding is used to model community formation in social networks \cite{boguna2004models, rodgers2024fitness, fernanda2023generaltraitecology}. For instance, communities are formed in co-author networks as researchers with similar research interests collaborate more often \cite{newman2004coauthorship,newman2001scientific,barabasi2002evolution}. Also, in protein interaction networks, proteins that are responsible for the functioning of similar biochemical pathways or have similar roles group into communities \cite{peregrin2009modular}. Furthermore, a larger group of relatively similar nodes can divide into smaller groups containing nodes with more specific similarities \cite{simon1962architecture,schaub2023hierarchical,clauset2008hierarchical}. 

Clearly, the variations in local properties of the nodes affect the organization at different scales in the network. We investigate this relationship through a phenomenological model, where two nodes with similar inherent properties connect with higher probability. Using this approach, we replicate the scaling relations that we discover are universal across multiple real-world networks. These scaling laws are derived from representing the hierarchical communities in a tree representation \cite{guimera2003self} in analogy to river networks \cite{horton1945erosional,scheidegger1968horton}. Further, we explain that such universal scaling relations arise due to a general self-organizing principle that allows ordered structures (communities) to emerge across different organizational scales of the network. According to Haken \cite{haken2012advanced}, the entropy (degree of freedom) of a system decreases when order emerges. We show that for a hierarchical organization to emerge, the diversity of the inherent properties of nodes within communities is minimized at every organizational scale of the network. We answer the following questions: (i) What are the universal features of hierarchical community structures found in diverse real-world networks? (ii) How does such universality emerge from local interactions among nodes? (iii) What is the general self-organizing principle that links the local interactions to the global emergent structure?


\section*{Universal scaling laws of hierarchical communities in real-world complex networks}

Organization of a network into communities-within-communities is evident in diverse systems. To describe the topology of such networks, we identify communities at several organizational scales using Girvan-Newman's algorithm \cite{girvan2002community} (see details in Materials and Methods). Using this algorithm, we start by classifying the nodes in the network into two prominent communities at the largest scale. Subsequently, for each of the large-scale communities we identify two prominent sub-communities and repeat this process until we break down the network to the smallest possible communities, i.e., the individual nodes in the network. In summary, we reveal a topology where the network bifurcates into communities which further bifurcate into sub-communities and so on. We map this hierarchical structure of communities onto a binary tree representation \cite{guimera2003self} where the network is at the top of the hierarchy (refer Materials and Methods). A community in the network is represented by a node in the tree. The bifurcation of a community into constituent sub-communities in the network is represented by branches connecting a parent community-node to children community-nodes in the tree.

\begin{figure}
    \centering
    \includegraphics[width=0.8\textwidth]{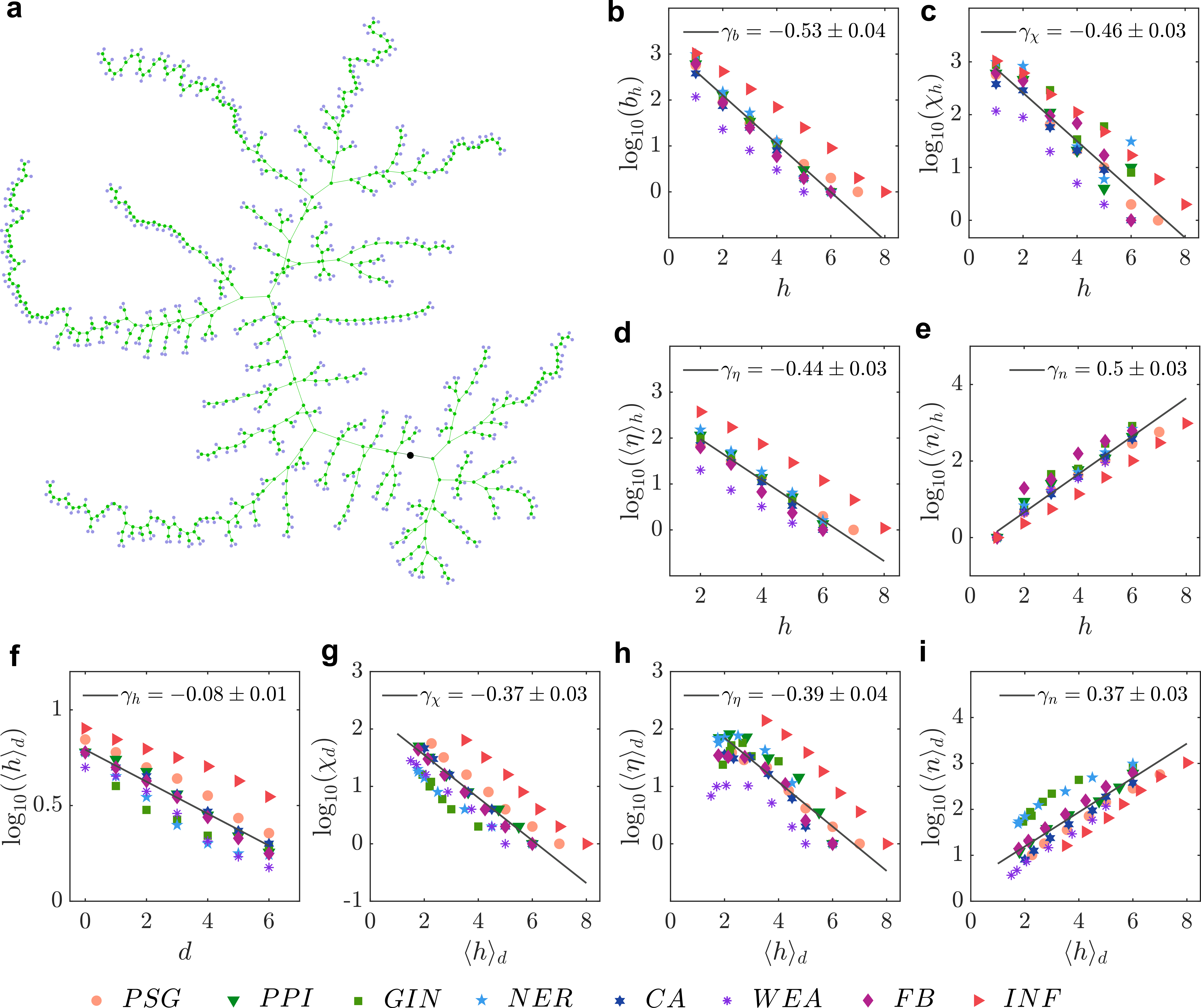}
    \caption{\textbf{Hierarchical community structure of real-world networks is structurally Hortonian and hence, topologically self-similar.} \textbf{(a)} Binary tree representation of the protein-protein interaction network of bacterium \textit{Escherichia coli} in figure \ref{fig_realntwk}(a). The network is represented as a big black colored node in the binary tree representation. \textbf{(b)} Variation of the logarithm of number of branches $b_h$ having the same order $h$ with $h$ for different real-world networks. Variation of the logarithm of \textbf{(c)} number of communities $\chi_h$, \textbf{(d)} mean relative link density of communities $\langle \eta \rangle _h$ and \textbf{(e)} mean size of communities $\langle n \rangle _h$ having the same index $h$ with respect to $h$. \textbf{(f)} Variation of the logarithm of mean order of communities $\langle h \rangle _d$ at fixed hierarchical depth $d$ with $d$. Variation of the logarithm of \textbf{(g)} number of communities $\chi_d$, \textbf{(h)} mean relative link density of communities $\langle \eta \rangle _d$ and \textbf{(i)} mean size of communities $\langle n \rangle _d$ with the mean index of communities $\langle h \rangle _d$. The subscript $d$ implies that the communities are taken at a fixed hierarchical depth $d$. Abbreviations of real-world networks are PSG: protein structure graph of a protein complex, PPI: protein-protein interactions in bacterium \textit{Escherichia coli}, GIN: gene interaction network of the worm \textit{Caenorhabditis elegans}, NER: network of nerve fibre tracts in mouse, CA: coauthorship network of scientists working on network science, WEA: interaction network of weavers, FB: network of mutually liked Facebook pages, and INF: infrastructure network of roads connecting cities in Europe.
    }
    \label{fig_HA_realntwk}
\end{figure}

The tree representation can depict the topological self-similarity between the composition of communities at multiple organizational scales and that of the entire network (e.g., see the self-similar tree structure in Fig. \ref{fig_HA_realntwk}(a)). Next, we quantify the self-similarity in the branching structure of the tree representation using the Horton-Strahler indexing scheme \cite{horton1945erosional, strahler1952dynamic, guimera2003self, arenas2004community}. This scheme assigns the index $h$ to a community-node in the tree based on the organizational scale of the corresponding community in the network. The smallest scale communities have the smallest organizational scale and are assigned $h=1$. When two communities of the same organizational scale (say $h$) constitute a larger community, then the organizational scale of the larger community is increased to $h+1$ (see Materials and Methods for details on the indexing scheme). 

For a complex network with self-similar topology of hierarchical communities, the corresponding tree representation has a self-similar branching structure \cite{guimera2003self}. Such a structure entails a geometric progression between the number of branches $b_h$ having indices $h$ \cite{horton1945erosional}. Hence, $R_b={b_h}/{b_{h+1}}=\text{constant}$, which implies an exponential scaling of $b_h$ versus $h$ known as the Horton's law of branch numbers: $\log_{10}{b_h} = \gamma_b h + c$, where $\gamma_b = -\log_{10}{R_b}$. 

\def\realgb{$-0.53\pm0.04$}
\def\realgx{$-0.46\pm0.03$}
\def\realge{$-0.44\pm0.03$}
\def\realgn{$0.50\pm0.03$}
\def\realgh{$-0.08\pm0.01$}
\def\realgxd{$-0.37\pm0.03$}
\def\realged{$-0.39\pm0.04$}
\def\realgnd{$0.37\pm0.03$}
\def\smgb{$-0.51 (1.00)$}
\def\smgx{$-0.45 (0.98)$}
\def\smge{$-0.42 (1.00)$}
\def\smgn{$0.50 (1.00)$}
\def\smgh{$-0.07 (0.98)$}
\def\smgxd{$-0.38 (1.00)$}
\def\smged{$-0.42 (0.98)$}
\def\smgnd{$0.38 (1.00)$}
\def\psggb{$-0.44 (0.97)$}
\def\psggx{$-0.49 (0.98)$}
\def\psgge{$-0.37 (1.00)$}
\def\psggn{$0.45 (0.99)$}
\def\psggh{$-0.08 (0.99)$}
\def\psggxd{$-0.37 (0.99)$}
\def\psgged{$-0.35 (0.99)$}
\def\psggnd{$0.37 (0.99)$}
\def\ppigb{$-0.55 (1.00)$}
\def\ppigx{$-0.45 (0.88)$}
\def\ppige{$-0.48 (1.00)$}
\def\ppign{$0.48 (0.96)$}
\def\ppigh{$-0.09 (0.98)$}
\def\ppigxd{$-0.38 (0.99)$}
\def\ppiged{$-0.43 (0.90)$}
\def\ppignd{$0.38 (0.99)$}
\def\gingb{$-0.59 (0.99)$}
\def\gingx{$-0.40 (0.90)$}
\def\ginge{$-0.48 (1.00)$}
\def\gingn{$0.57 (0.97)$}
\def\gingh{$-0.08 (0.89)$}
\def\gingxd{$-0.29 (0.91)$}
\def\ginged{$-0.38 (0.78)$}
\def\gingnd{$0.29 (0.91)$}
\def\nergb{$-0.61 (0.99)$}
\def\nergx{$-0.41 (0.77)$}
\def\nerge{$-0.48 (1.00)$}
\def\nergn{$0.54 (0.99)$}
\def\nergh{$-0.09 (0.95)$}
\def\nergxd{$-0.31 (0.98)$}
\def\nerged{$-0.40 (0.88)$}
\def\nergnd{$0.31 (0.98)$}
\def\cagb{$-0.52 (0.99)$}
\def\cagx{$-0.51 (0.96)$}
\def\cage{$-0.46 (0.99)$}
\def\cagn{$0.50 (1.00)$}
\def\cagh{$-0.08 (0.99)$}
\def\cagxd{$-0.42 (0.99)$}
\def\caged{$-0.42 (0.94)$}
\def\cagnd{$0.42 (0.99)$}
\def\weagb{$-0.50 (0.99)$}
\def\weagx{$-0.48 (0.97)$}
\def\weage{$-0.38 (1.00)$}
\def\weagn{$0.48 (0.98)$}
\def\weagh{$-0.10 (0.98)$}
\def\weagxd{$-0.40 (1.00)$}
\def\weaged{$-0.25 (0.77)$}
\def\weagnd{$0.41 (1.00)$}
\def\fbgb{$-0.56 (0.98)$}
\def\fbgx{$-0.52 (0.91)$}
\def\fbge{$-0.47 (0.99)$}
\def\fbgn{$0.52 (0.92)$}
\def\fbgh{$-0.09 (0.99)$}
\def\fbgxd{$-0.39 (1.00)$}
\def\fbged{$-0.38 (0.91)$}
\def\fbgnd{$0.39 (1.00)$}
\def\infgb{$-0.44 (0.99)$}
\def\infgx{$-0.39 (0.99)$}
\def\infge{$-0.41 (0.99)$}
\def\infgn{$0.42 (1.00)$}
\def\infgh{$-0.06 (0.99)$}
\def\infgxd{$-0.42 (0.99)$}
\def\infged{$-0.49 (0.99)$}
\def\infgnd{$0.42 (0.99)$}

Here, we analyze the network structures derived from multiple real-world systems including the protein structure graph of a protein complex, protein-protein interactions in bacterium \textit{Escherichia coli}, gene interaction network of the worm \textit{Caenorhabditis elegans}, network of nerve fibre tracts in mouse, co-authorship network of scientists working on network science, ecological interactions among weavers, network of mutually liked Facebook pages, and infrastructure network of roads connecting cities in Europe.

We discover that the exponent $\gamma_b$ of the semi-log scaling between $b_h$ and $h$ is strikingly similar for the tree representations of manifold real-world networks (Fig. \ref{fig_HA_realntwk}(b)). That is, several real-world systems exhibit similar topology of nested communities. We find that $\gamma_b \approx -0.53$ (with a standard error of $\pm 0.04$, $>90\%$ confidence) corresponding to a bifurcation ratio $R_b \approx 3.38(\pm0.01)$. The Horton-Strahler indexing scheme was originally introduced for quantifying the self-similarity in river systems. Horton discovered that the bifurcation ratio for rivers branching into smaller rivers and brooks was $R_b \approx 3.5$ across several river basins. Such striking universality across river basins as well as in networks derived from diverse real-world complex systems incites the idea that the emergence of hierarchical organizations are related to general underlying principles. 

Furthermore, self-similar trees are known to exhibit scaling relations between $h$ and the mean attributes along the branches of the same $h$ in the tree \cite{kovchegov2022random}. Here, we explore such relations for the mean attributes of community-nodes in the tree representations of complex networks. We define for fixed $h$, (a) $\chi_{h}$: the number of communities, (b) $\langle d\rangle_h$: the mean hierarchical depth, (c) $\langle n \rangle_{h}$: the mean size and (d) $\langle \eta \rangle_{h}$: the mean of the relative link density of communities; the mean is calculated across communities with same index $h$. The size of a community $C_j$ is the number of nodes in the community, denoted by $n_{C_j}$. The relative link density $\eta_{C_j}$ is the ratio of the link density of the community ($\rho_{C_j}$) with respect to that of the entire network ($\rho_{C_0}$). For a community $C_j$ comprising $n_{C_j}$ nodes, the link density is defined as $\rho_{C_j} = \frac{\sum _{i \in C_j} k_{i}^{C_j}/2}{n_{C_j}(n_{C_j}-1)/2}$. Here, $k_{i}^{C_j}$ is the intra-community degree, i.e., the number of connections between a node $i$ in community $C_j$ with other nodes in the same community. At the smallest scales, where individual nodes constitute separate communities, the relative link density is set to zero. Also, if $\eta_{C_j} > 1$, we infer that the nodes within the community $C_j$ are more densely connected than the whole network. 

Interestingly, we find that the quantity $\chi_{h}$ and the mean attributes ($\langle \eta \rangle_{h}$, $\langle n \rangle_{h}$) of communities with the same order $h$ exhibit (i) unique scaling relations with the order $h$, and (ii) striking similarity in these scaling relations across diverse real-world networks (see Fig. \ref{fig_HA_realntwk}(c-e) and refer Tables \ref{table_univLaw_realNmodel} and \ref{table_realworld}). 
These unique scaling relations imply that, not only the topology, but also the pattern of connections within communities is self-similar across various organizational scales. Moreover, such similarity in the composition of communities across hierarchical scales is described by universal scaling laws across diverse systems.

\begin{table}[h]
\centering
\begin{tabular}{|c|c|r|r|}
\hline
\multirow{2}{*}{Coefficient} & \multirow{2}{*}{$y$ versus $x$}                                & \multicolumn{1}{c|}{\multirow{2}{*}{\begin{tabular}[c]{@{}c@{}}Real-world\\ networks\end{tabular}}} & \multicolumn{1}{c|}{\multirow{2}{*}{\begin{tabular}[c]{@{}c@{}}Status model\\ network\end{tabular}}} \\
                             &                                                                & \multicolumn{1}{c|}{}                                                                               & \multicolumn{1}{c|}{}                                                                                \\ \hline
$\gamma_b$                   & $\log_{10}(b_h)$ vs $h$                                        & \realgb                                                                                             & \smgb                                                                                                \\ \hline
$\gamma_\chi$                & $\log_{10}(\chi_h)$ vs $h$                                     & \realgx                                                                                             & \smgx                                                                                                \\ \hline
$\gamma_\eta$                & $\log_{10}(\langle \eta \rangle _h)$ vs $h$                    & \realge                                                                                             & \smge                                                                                                \\ \hline
$\gamma_n$                   & $\log_{10}(\langle n \rangle _h)$ vs $h$                       & \realgn                                                                                             & \smgn                                                                                                \\ \hline
$\gamma_{h}$                 & $\log_{10}(\langle h \rangle _d)$ vs $d$                       & \realgh                                                                                             & \smgh                                                                                                \\ \hline
$\gamma_{\chi_d}$            & $\log_{10}(\chi _d)$ vs $\langle h \rangle _d$                 & \realgxd                                                                                            & \smgxd                                                                                               \\ \hline
$\gamma_{\eta_d}$            & $\log_{10}(\langle \eta \rangle _d)$ vs $\langle h \rangle _d$ & \realged                                                                                            & \smged                                                                                               \\ \hline
$\gamma_{n_d}$               & $\log_{10}(\langle n \rangle _d)$ vs $\langle h \rangle _d$    & \realgnd                                                                                            & \smgnd                                                                                               \\ \hline
\end{tabular}
\caption{A comparison between the Horton scaling exponents observed for different real-world networks and the network obtained from the model. The scaling exponents of different relations are reported for real-world networks (mean of exponents in Table \ref{table_realworld} with $90\%$ confidence) and for the network obtained from the model network (with the goodness of fit (R-Square) in brackets).}
\label{table_univLaw_realNmodel}
\end{table}

\begin{table}[h]
\centering
\resizebox{\columnwidth}{!}{%
\begin{tabular}{|c|c|rrrrrrrr|}
\hline
\multirow{2}{*}{Coefficient} & \multirow{2}{*}{$y$ versus $x$}                                & \multicolumn{8}{c|}{Real-world networks}                                                                                                                                                                                                        \\ \cline{3-10} 
                             &                                                                & \multicolumn{1}{c|}{PSG}     & \multicolumn{1}{c|}{PPI}     & \multicolumn{1}{c|}{GIN}     & \multicolumn{1}{c|}{NER}     & \multicolumn{1}{c|}{CA}     & \multicolumn{1}{c|}{WEA}     & \multicolumn{1}{l|}{FB}     & \multicolumn{1}{l|}{INF} \\ \hline
$\gamma_b$                   & $\log_{10}(b_h)$ vs $h$                                        & \multicolumn{1}{r|}{\psggb}  & \multicolumn{1}{r|}{\ppigb}  & \multicolumn{1}{r|}{\gingb}  & \multicolumn{1}{r|}{\nergb}  & \multicolumn{1}{r|}{\cagb}  & \multicolumn{1}{r|}{\weagb}  & \multicolumn{1}{r|}{\fbgb}  & \infgb                   \\ \hline
$\gamma_\chi$                & $\log_{10}(\chi_h)$ vs $h$                                     & \multicolumn{1}{r|}{\psggx}  & \multicolumn{1}{r|}{\ppigx}  & \multicolumn{1}{r|}{\gingx}  & \multicolumn{1}{r|}{\nergx}  & \multicolumn{1}{r|}{\cagx}  & \multicolumn{1}{r|}{\weagx}  & \multicolumn{1}{r|}{\fbgx}  & \infgx                   \\ \hline
$\gamma_\eta$                & $\log_{10}(\langle \eta \rangle _h)$ vs $h$                    & \multicolumn{1}{r|}{\psgge}  & \multicolumn{1}{r|}{\ppige}  & \multicolumn{1}{r|}{\ginge}  & \multicolumn{1}{r|}{\nerge}  & \multicolumn{1}{r|}{\cage}  & \multicolumn{1}{r|}{\weage}  & \multicolumn{1}{r|}{\fbge}  & \infge                   \\ \hline
$\gamma_n$                   & $\log_{10}(\langle n \rangle _h)$ vs $h$                       & \multicolumn{1}{r|}{\psggn}  & \multicolumn{1}{r|}{\ppign}  & \multicolumn{1}{r|}{\gingn}  & \multicolumn{1}{r|}{\nergn}  & \multicolumn{1}{r|}{\cagn}  & \multicolumn{1}{r|}{\weagn}  & \multicolumn{1}{r|}{\fbgn}  & \infgn                   \\ \hline
$\gamma_{h}$                 & $\log_{10}(\langle h \rangle _d)$ vs $d$                       & \multicolumn{1}{r|}{\psggh}  & \multicolumn{1}{r|}{\ppigh}  & \multicolumn{1}{r|}{\gingh}  & \multicolumn{1}{r|}{\nergh}  & \multicolumn{1}{r|}{\cagh}  & \multicolumn{1}{r|}{\weagh}  & \multicolumn{1}{r|}{\fbgh}  & \infgh                   \\ \hline
$\gamma_{\chi_d}$            & $\log_{10}(\chi _d)$ vs $\langle h \rangle _d$                 & \multicolumn{1}{r|}{\psggxd} & \multicolumn{1}{r|}{\ppigxd} & \multicolumn{1}{r|}{\gingxd} & \multicolumn{1}{r|}{\nergxd} & \multicolumn{1}{r|}{\cagxd} & \multicolumn{1}{r|}{\weagxd} & \multicolumn{1}{r|}{\fbgxd} & \infgxd                  \\ \hline
$\gamma_{\eta_d}$            & $\log_{10}(\langle \eta \rangle _d)$ vs $\langle h \rangle _d$ & \multicolumn{1}{r|}{\psgged} & \multicolumn{1}{r|}{\ppiged} & \multicolumn{1}{r|}{\ginged} & \multicolumn{1}{r|}{\nerged} & \multicolumn{1}{r|}{\caged} & \multicolumn{1}{r|}{\weaged} & \multicolumn{1}{r|}{\fbged} & \infged                  \\ \hline
$\gamma_{n_d}$               & $\log_{10}(\langle n \rangle _d)$ vs $\langle h \rangle _d$    & \multicolumn{1}{r|}{\psggnd} & \multicolumn{1}{r|}{\ppignd} & \multicolumn{1}{r|}{\gingnd} & \multicolumn{1}{r|}{\nergnd} & \multicolumn{1}{r|}{\cagnd} & \multicolumn{1}{r|}{\weagnd} & \multicolumn{1}{r|}{\fbgnd} & \infgnd                  \\ \hline
\end{tabular}%
}
\caption{Horton scaling exponents observed for different real-world networks. The scaling exponents are listed with the goodness of fit (R-Square) in brackets for different real-world networks.}
\label{table_realworld}
\end{table}

Importantly, we note that a self-similar binary tree exhibits Horton scaling relations; however, the reverse is not necessarily true \cite{scheidegger1968horton}. A tree that exhibits Horton scaling relations need not be structurally self-similar \cite{scheidegger1968horton}. \textit{Structural self-similarity} implies that a sub-tree has similar bifurcation and structural properties as the whole tree. To quantify structural self-similarity, we identify a subset of communities that are nodes in the tree at a fixed hierarchical depth. At a fixed depth, communities of several organizational scales can be present. However, as the hierarchical depth increases, the range of $h$ and the mean organizational scale of the communities decreases. We define $\langle h \rangle_d$, the mean organizational scale at a fixed hierarchical depth $d$ as the mean of the Horton-Strahler index of all communities at that depth. Figure \ref{fig_HA_realntwk}(f) shows that $\langle h \rangle_{d}$ decreases exponentially with the hierarchical depth $d$. Also, this scaling relation is universal across multiple real-world networks and implies universality in the organizational structure of sub-trees at different hierarchical depths.

Similarly, we define $\langle \chi \rangle_{d}$, $\langle \eta \rangle_{d}$, $\langle n \rangle_{d}$ as mean attributes of communities at fixed depth. For a structurally self-similar tree, the variation of mean attributes of communities with the mean organizational scale ($\langle h \rangle_{d}$) must be self-similar across various hierarchical depths. We find that, $\langle \chi \rangle_{d}$, $\langle \eta \rangle_{d}$, $\langle n \rangle_{d}$ follow distinct scaling relations with $\langle h \rangle_d$ (see Fig. \ref{fig_HA_realntwk}(g-i)). Moreover, we uncover that the scaling relations in Fig. \ref{fig_HA_realntwk}(g-i) are universal across the multiple real-world networks. Note that, community structures may remain prominent across only within a certain range of $d$. Hence, the scaling relations are obeyed for a certain range of hierarchical depths starting from $d = 0$. The scaling relations shown in Fig. \ref{fig_HA_realntwk}(g-i) are shown for communities with depth in the range $d\in[0,6]$. Note that, such scaling relations obtained at fixed $d$ vanish for a tree that is not structurally self-similar, even if the tree exhibits Horton's law of branch numbers (see Supplementary material S1 for details). 

In summary, we discover that the community-within-community structure is \textit{structurally self-similar} described by scaling laws that are universal across diverse real-world complex systems.   


\section*{Emergence of topological self-similarity through local link formation rules}

Universality in the emerging patterns of a self-similar hierarchical community structure implies that, a general mechanism for such emergence exists independent of the specific details of the system. In a complex system, explaining the emergence of global patterns and structures while accounting for local interactions is a major challenge. Universal emergent features incite us to look for similarities in local interactions across different systems. Communities are formed when nodes form groups based on their similarities \cite{newman2003structure,fortunato2010community,newman2012communities, schleussner2016clustered}. This understanding has been exploited to explain the formation of hierarchical communities in social networks \cite{boguna2004models, rodgers2024fitness}. 

\begin{figure}
    \centering
    \includegraphics[width=0.7\textwidth]{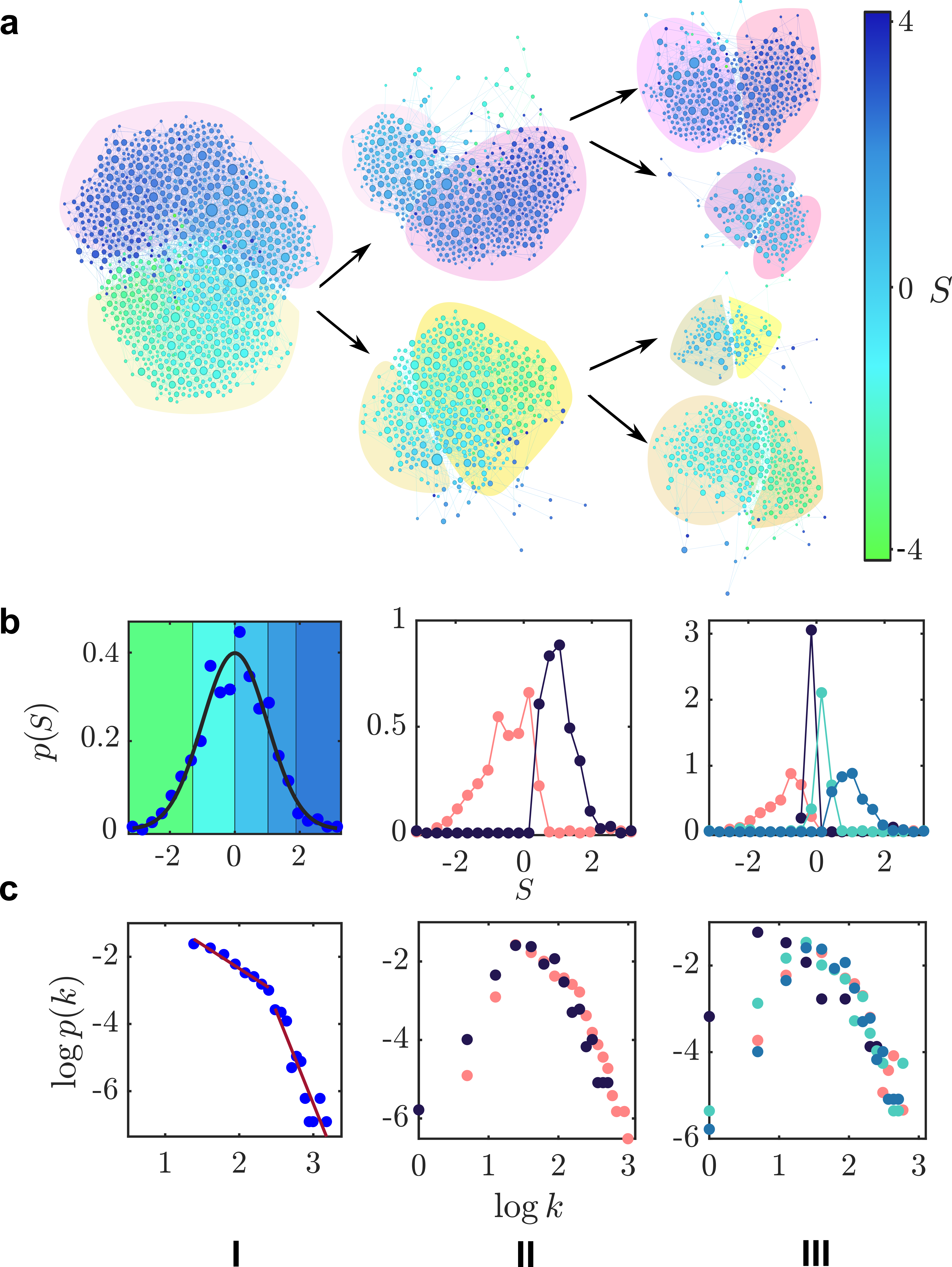}
    \caption{\textbf{Similar nodes are more likely to connect and group together to form communities at multiple scales.} \textbf{(a)} An illustration of the community-within-community structure in a network obtained from the model (visualized using Gephi \cite{bastian2009gephi}). The network is visualized in a force-directed layout that spreads the nodes spatially into communities. The nodes are colored based on their statuses $S$. The network is simulated using the input parameters: $N = 1000$, $m = 4$, $N_0 = 100$ and $p(S)$ is a Gaussian probability density function with zero mean and unit variance. \textbf{(b)} Probability distribution of statuses ($p(S)$ versus $S$) of all the nodes in (I) the network fitted by a Gaussian status distribution (black curve) and (II, III) for communities detected at subsequent scales of organization. The colored boxes in the background of the plot in (b-I) illustrate the heterogeneity in the status distribution. \textbf{(c)} Degree distribution ($p(k)$ versus $k$) in log-log scale for (I) the network and (II, III) communities at subsequent scales of organization. Here, $k$ denotes the number of intra-community connections of a node.}
    \label{fig_status_model}
\end{figure}

We translate this understanding to a fundamental rule of local link formation in our model for network construction: nodes that are more similar to each other are more likely to form links. To each node $i$, we assign a value referred as \textit{status} $S_i$ representing an intrinsic property of that node. Statuses of nodes are derived from a non-uniform probability density function, referred as the status distribution and denoted as $p(S)$. Here, we use a Gaussian probability distribution with zero mean and unit variance (see Fig. \ref{fig_status_model}(b-I)). Starting from a set of randomly connected $N_0$ nodes, the model evolves to a total of $N$ nodes. Thus, the probability $\pi _{ij}$ that an incoming node $i$ forms a connection with an existing node $j$ in the network is given by Eq. \ref{eq_probab}. 
\begin{equation}\label{eq_probab}
    \pi _{ij} = \frac{d_{ij}}{\sum_k d_{ik}}    
\end{equation}
Here, $d_{ij}$ is the inverse of the difference in the statuses of two nodes $i$ and $j$, i.e., $d_{ij}={1}/|{S_i - S_j}|$. The higher the difference between the intrinsic properties of the two nodes, the lower the probability of them forming a connection. This probability function accounts for \textit{both} nodes involved in the link formation, unlike previous approaches \cite{erdos1960evolution,watts1998collective,barabasi1999emergence,bianconi2001bose}. Also, contrary to models of preferential attachment that are based on the number of connections (degree) of a node \cite{bianconi2001bose, dorogovtsev2002evolution, rodgers2024fitness}, the probability of link formation here is independent of the degree of nodes. We regard the degree of a node rather as a result of, than as a factor for, link formation. Every node in the network may not always have the information about the connectivity of all other nodes. Moreover, any effect of degree in determining the probability of link formation can be assumed to be reflected in the `status' of the node. 



\begin{figure}
    \centering
    \includegraphics[width=0.8\textwidth]{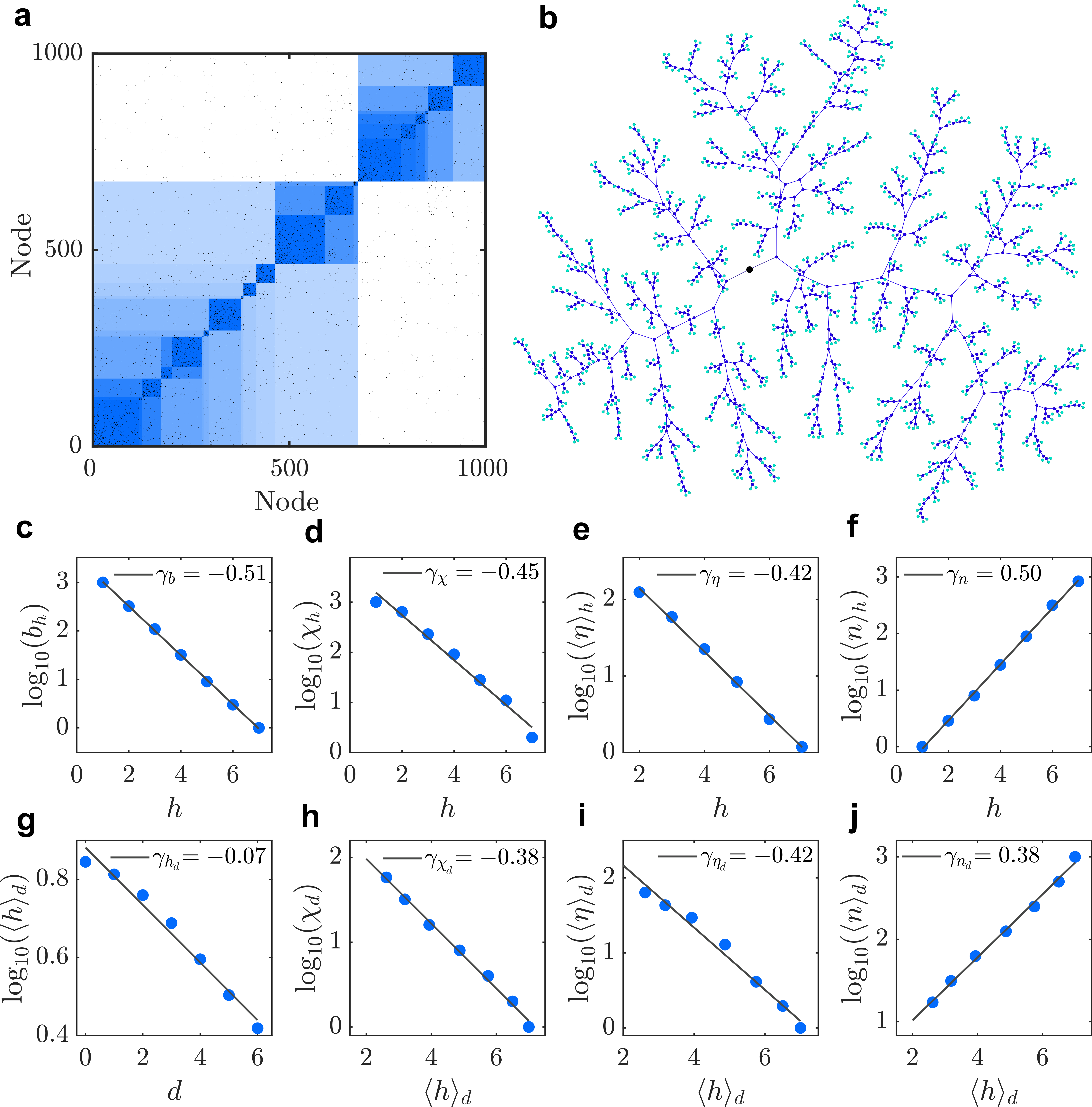}
    \caption{\textbf{Network obtained from our model exhibits a highly self-similar topology with a structurally self-similar hierarchical community structure.} \textbf{(a)} The adjacency matrix of the network obtained from the model. The colored blocks highlight communities and the opacity of color increases with the decreasing scale of organization. \textbf{(b)} Binary tree representation of the network in (a) with an evidently self-similar branching structure. \textbf{(c)} Plot of the logarithm of number of branches $b_h$ having the same index $h$. Plot of the logarithm of \textbf{(d)} number of communities $\chi_h$, \textbf{(e)} mean relative link density of communities $\langle \eta \rangle _h$ and \textbf{(f)} mean size of communities $\langle n \rangle _h$ having the same index $h$. \textbf{(g)} Variation of logarithm of mean order  $\langle h \rangle_d$ at fixed hierarchical depth $d$ with $d$. Plot of the logarithm of \textbf{(h)} number of communities $\chi_d$, \textbf{(i)} mean relative link density of communities $\langle \eta \rangle _d$ and \textbf{(j)} mean size of communities $\langle n \rangle _d$ with the mean index $\langle h \rangle_d$. The subscript $d$ denotes that the communities considered are at a fixed hierarchical depth $d$. Refer Table \ref{table_univLaw_realNmodel} for a comparison of the scaling exponents obtained from the model and the real-world networks.}
    \label{fig_HA_status_model}
\end{figure}

We find that a complex network with self-similar hierarchical community structure emerges from the rule of local interactions governed by Eq. \ref{eq_probab}; see Fig. \ref{fig_status_model}(a) and the corresponding tree representation in Fig. \ref{fig_HA_status_model}(b). This network exhibits the same universal scaling relations (shown in Fig. \ref{fig_HA_status_model}) as those observed across multiple real-world networks. The effect of distinct status distributions $p(S)$, and different values of $m$ and $N$ on the network topology are discussed in detail in the Supplementary material S2-S4. Also note, a version of the model, where the network does not grow with time and nodes form connections based on the similarity of their statuses, is discussed in the Supplementary material S4; growth in the model is not necessary to replicate the scaling laws observed in real-world networks. 

Using community detection \cite{newman2004detecting}, we find that the nodes group primarily into two large-scale communities. Within each community, nodes with relatively more similarities in their statuses further regroup into smaller tight-knit communities. The differentiation of nodes within a community into sub-communities occurs such that the status distributions of the sub-communities are almost non-overlapping; e.g. see the status distribution of communities formed at different scales in Fig. \ref{fig_status_model}(b-II,III).

The degree distribution of the network obtained from the model (Fig. \ref{fig_status_model}(c)) exhibits a power-law like variation, where the scaling exponent varies with the range of degree, similar to that in scientific collaboration networks \cite{newman2001scientific}. For high values of degree ($k$), the intra-community degree distribution of communities (Fig. \ref{fig_status_model}(c-II,III)) appear similar to that of the network; this indicates a similarity in the pattern of connectivity between the network and communities formed at distinct scales. However, for low values of $k$, we find significant deviation from a power-law like behavior in Fig. \ref{fig_status_model}(c-II,III). The pattern of connectivity of the network is further visualized using an adjacency matrix shown in Fig. \ref{fig_HA_status_model}(a). Note the diagonal blocks-within-blocks pattern in the matrix with each block representing a community. The embedding of smaller blocks within the larger blocks appears self-similar.

Interestingly, the tree representation (Fig. \ref{fig_HA_status_model}(b)) of the network obtained from the model delineates a highly self-similar branching structure and exhibits Horton's laws of branch numbers (Fig. \ref{fig_HA_status_model}(c)) and mean attributes (Fig. \ref{fig_HA_status_model}(d-f)). The scaling exponents obtained from the self-similar topology of the network simulated from the model are intriguingly close to the scaling exponents obtained from various real-world networks and are listed in Table \ref{table_univLaw_realNmodel}. Moreover, the model produces a network with \textit{structurally self-similar} hierarchical communities with scaling exponents similar to those of diverse real-world complex networks (see Fig. \ref{fig_HA_status_model}(g-j) and Table \ref{table_univLaw_realNmodel}). The structural self-similarity of hierarchical communities arising from such a basic model is intriguing. The network is formed due to local interactions based on the mutual similarity of nodes. The nature of these local interactions leads to the formation of organised groups (communities) at multiple scales. Hence, this approach can be used as a generating mechanism for simulating networks with self-similar hierarchical communities.



\section*{A general self-organizing principle results in universality in hierarchical community structure}

We argue that the community-within-community structure emerges not only from local interactions between nodes but also due to the self-organization of nodes at many `scales'. What does the process of self-organization entail and why does such organization lead to a self-similar hierarchical community structure? In the model, the structure of the emergent network divides the nodes with diverse statuses into separate communities; each community divides into sub-communities due to further differentiation of statuses within a smaller range of the status distribution. Thus, communities are formed at multiple scales due to the possibility of differentiation between nodes in multiple ranges of the status distribution. Here, we show that the emergent structure of the network optimizes the relative similarity of nodes in order to form communities at multiple organizational scales.

\begin{figure}
    \centering
    \includegraphics[width=1\textwidth]{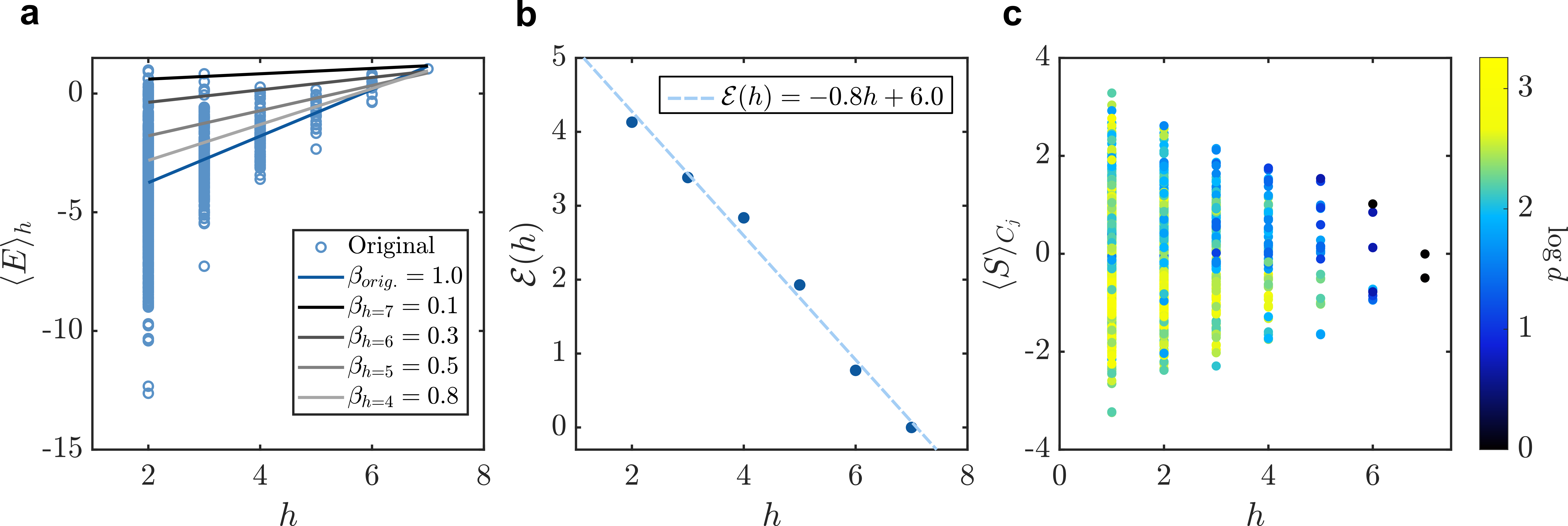}
    \caption{\textbf{Self-organization leads to a community-within-community structure that minimizes the diversity among nodes at each organizational scale in the network.} (a) Variation with $h$ of the mean entropy $\langle E \rangle_h$ representing the diversity of statuses of nodes within communities of order $h$. The entropy of a community $C_j$ of order $h$ is represented by a blue circle marker ({$\circ$}), the mean entropy $\langle E \rangle_h$ is plotted using a blue solid line with slope $\beta=1.0$. The gray solid lines represent surrogate cases of the variation of $\langle E \rangle_h$ where the value of statuses are randomly permuted among nodes within each community detected at some fixed organizational scale $h_k$, but not across the communities. The slope $\beta_{h=h_k}$ of $\langle E \rangle_h$ vs $h$ decreases as the surrogate test is performed at increasing values of $h_k$; indicating that entropy is most optimally minimized at each scale by the original organization of nodes into communities of the network. (b) The variation of organizational entropy $\mathcal{E}(h)$ with the organizational scale $h$. (c) Distribution of the mean status $\langle S\rangle_{C_j}$ of communities with the organizational scale $h$. The mean status of communities diversifies with order $h$.}
    \label{fig_entropy_surrogate}
\end{figure}

To quantify the diversity among nodes within a community $C_j^h$ at an organizational scale $h$, we define entropy of a community as
\begin{equation}\label{eq_entropy_Cj}
    E_{C_j^h}(S)=-\int \limits_{S\in C_j^h} p(S) \text{ln}(p(S))dS
\end{equation}
Here, $E_{C_j^h}(S)$ is called the \textit{continuous entropy} \cite{renyi1961measures,conrad2004probability,marsh2013introduction} and is defined using $p(S)$, i.e., the probability density function of the statuses of nodes contained in the community $C_j^h$. The optimal bin width for computing the integral is determined using the Freedman–Diaconis rule \cite{freedman1981histogram}. We note that, although this integral form of entropy is analogous to the discrete form of Shannon entropy, $E_{C_j^h}(S)$ can assume negative values as well \cite{conrad2004probability, marsh2013introduction}. Yet, we can interpret $E_{C_j^h}(S)$ as a quantification of the diversity of statuses of nodes in the community $C_j^h$. The lower the value of $E_{C_j^h}(S)$, the lower the diversity among nodes in that community. As the scale $h$ decreases, the number of communities identified at that scale increases. Hence, the values of the entropy $E_{C_j}$ (Fig. \ref{fig_entropy_surrogate}(a), blue circles) are distributed in a wide range at lower values of $h$. 

The average continuous entropy of communities increases $\langle E \rangle_h$ with increase in $h$ (solid blue line in Fig. \ref{fig_entropy_surrogate}(a)) depicting that the diversity among nodes is greater in a large community identified at a higher organizational scale. Clearly, the structural organization of the nodes into hierarchical communities is related to the underlying status distribution. But, how optimal is the distribution of statuses within each community? 
To explore this relation, we compare the status entropy of communities in the original network with surrogate cases \cite{lancaster2018surrogate}, where the structure is preserved but the status distribution is randomized. That is, we ask, what would be the status entropy within each community across scales if the hierarchical community structure did not emerge due to the underlying status distribution, but rather by random chance (or stochastic probabilities of link formation). If indeed the network structure emerges from local interactions based on the node properties, then we expect that the average entropy of a surrogate hierarchical community structure will be higher than that of the original structure and the slope of $\langle E \rangle_h$ vs $h$ will increase.

For a surrogate case, we preserve the hierarchical community structure at all scales; however, we randomly permute the statuses of nodes within all communities identified at a fixed scale $h_k$. As a result, the status distributions within communities formed at scales $h\geq h_k$ are preserved, while the status distribution within the smaller communities formed at scale $h<h_k$ is randomized. We perform such surrogate analysis for different values of $h_k$, and define surrogate continuous entropy for communities at all scales. Specifically, we define $E_{surr}\{{C_j^h},h_k\}$ as the entropy of all the communities ($C_j^h$ identified at various scales $h$) when the random permutation of statuses is performed among nodes in communities at scale $h_k$. 

We find that, $\langle E_{surr} \rangle_h > \langle E \rangle_h$, for $h<h_k$, and this is true for all values of $h_k$. Hence, the slope $\beta_{h=h_k}$ of $\langle E_{surr} \rangle_h$ vs $h$ is lesser than that for the original status distribution of the network. Also, notice that this decrease in the slope $\beta_{h=h_k}$ is greater for higher values of $h_k$ (compare the slopes of solid gray lines in Fig. \ref{fig_entropy_surrogate}(a)). Clearly, the diversity (entropy) among nodes in a community identified at any arbitrary scale $h$ is minimum for the original network as compared to the surrogate cases with randomized status distributions. Thus, the emergent topology of the network is essentially the result of a self-organization among nodes that leads to the most optimal distribution of node properties within communities at each scale.

Furthermore, we define an organizational entropy $\mathcal{E}(h)$ as a function of the organizational scale $h$, as given in Eq. \ref{eq_Org_entropy}. First, we compute the difference between the continuous entropy of a community $C_j^h$ in the original network and the entropy of the same community when the statuses are randomly permuted throughout the network (i.e., for $h_k=\text{max}(h)$). Then, we find the mean of this quantity across communities at each organizational scale $h$. 

\begin{equation}\label{eq_Org_entropy}
    \mathcal{E}(h)=\langle \text{surr}\{E_{C_j^h},\text{max}(h)\}-  E_{C_j^h}(S) \rangle _h
\end{equation}
Note that, the absolute continuous entropy $E_{C_j}(S)$ of a community quantifies the composition (in terms of statuses of nodes) of that community alone. However, $\mathcal{E}(h)$ quantifies the \textit{orderliness} in the distribution of statuses across communities at each scale $h$ owing to the complex hierarchical community structure of the network. Thus, $\mathcal{E}(h)$ represents the information needed to describe the orderliness of the structure at a particular scale $h$. At the smallest organizational scales, where individual nodes constitute separate communities, there is negligible organization and we must describe each node separately (lot of information). As a result the $\mathcal{E}(h)$ is very high at lower values of $h$. However, at a higher value of $h$, we can identify distinct communities in the network and these communities represent the structure of the system at that scale. With increasing $h$, the network structure is represented by fewer and fewer communities. Thus, the information needed to describe the system at scale $h$ decreases with increase in $h$. As a result, the organizational entropy $\mathcal{E}(h)$ decreases with increasing scale $h$ (see Fig. \ref{fig_entropy_surrogate}(b)) and is zero at the largest organizational scale.

Finally, Fig. \ref{fig_entropy_surrogate}(c) shows that the mean statuses $\langle S\rangle_{C}$ of communities at each scale $h$ are distributed in a wide range. The mean status of the network is zero, same as the mean of the status distribution of nodes ($\mu=0$ for $p(S)$ in Fig. \ref{fig_status_model}(b)). Also, the distribution of the mean status of communities becomes wider at smaller values of $h$. The bell distribution of $\langle S\rangle_{C}$ versus $h$ for communities in Fig. \ref{fig_entropy_surrogate}(c) resembles the originally assumed Gaussian status distribution $p(S)$ for nodes. Clearly, the diversity in the inherent property of nodes reflects as diversity across communities. In Supplementary S2, we show that the distribution of mean statuses of communities are different for networks obtained from distinct underlying status distributions. However, we find that the self-organizing principle remains the same (as described through Fig. \ref{fig_entropy_surrogate}(a)) and hence, we observe striking universality in the scaling laws describing the topology of networks obtained from different status distributions as well (Supplementary S2).

Haken explained that entropy is a measure of disorder or the degree of freedom and variability, and entropy decreases in a self-organizing system \cite{haken2012advanced}. Here, we find that the diversity of statuses of nodes is minimized within communities at \textit{every scale of organization}. Thus, a network constructed from relative similarity of nodes self-organizes at multiple `scales'. Now, the definition of organizational scales becomes clear; an organizational scale is one at which tightly-knit groups of nodes emerge such that the diversity among the nodes within each community is minimized (compared to that in a similar structure formed by random chance). 

Moreover, as we increase our observational scale from the level of individual nodes to the entire network, we find that the complexity of the organization increases and the organizational entropy decreases. Note that, the organizational entropy decreases almost linearly with an increase in $h$ depicting that the Horton-Strahler order ($h$) is a robust index for quantifying the organizational scales of the network across which the self-organizing principle remains the same. Since the self-organizing principle is the same at all scales of organization, we find the emergence of universal scaling laws describing the self-similarity in the topological structure of such networks.



\section*{Discussion}

We study the formation of hierarchical communities in diverse systems spanning across social, infrastructural, biological and animal interaction networks. We discover that the emergent topology of communities-within-communities in these systems are self-similar and are described by unique scaling laws. These scaling laws are obtained by identifying communities formed at different organizational scales (order $h$). Universality of these scaling relations across systems implies that there is a general self-organizing principle that is independent of the details of the system. Using a basic model for network construction, we explain that self-similar hierarchical communities emerge when nodes with greater similarity are allowed to connect with higher probability. The universal scaling relations are recovered in this model by virtue of self-organization among the nodes. 

The emergence of ordered structures and patterns in real-world systems has always intrigued researchers. Contrary to intuition, several real-world systems evolve to form organized structures instead of portraying disorderliness. Emergence occurs when a system is complex and locally interacting components can self-organize into a pattern that emerges at a much larger scales. But why and how do the constituent entities of a complex system organize themselves? What benefits do individual entities accrue from participating in such an organization process?

We discover through our model that the process of self-organization is nothing but a process by which the diversity among nodes is minimized within the communities identified at distinct organizational scales. This understanding concurs with Haken's principle that entropy decreases for self-organizing systems. We show that the entropy of node properties is minimized at all organizational scales through the formation of hierarchical communities. Self-similarity in the emergent structure is intriguing as it indicates that the same process occurs at multiple scales. Moreover, the information needed to describe the structure, decreases from describing several nodes to communities to a whole network; hence, the `organizational entropy' decreases as the observational scale increases.

Our analyses advance the current understanding of why and how do locally interacting components self-organize in a complex system. Connections between nodes are sustained depending on the cost of maintaining a link between nodes. The higher the differences in the inherent properties of two nodes, the higher would be the cost of maintaining the link. Thus, minimizing the diversity among its neighbors is beneficial to each node, as well as to each sub-community within a larger community. The emergence of an ordered structure is beneficial to each constituent entity, to various functional groups and as well as to the functioning of the whole system. Interestingly, such optimization at the level of nodes (local) and the network (global) occurs \textit{spontaneously} without any special constraints applied in our model. 

Further, our findings incite novel insights about specific real-world systems as well. For example, consider a scientific collaboration network. It is well known that researchers with similar interests or in similar disciplines come together to form communities in the collaboration network. Researchers also collaborate transcending scientific disciplines often leading to novel discoveries (inter-community connections); however, the cost of communication among researchers from different fields or academic training is greater as compared to that among researchers from similar backgrounds. Hence, we find tightly knit communities of researchers in similar disciplines that evolve perhaps to reduce the cost of scientific communication. Then, we can infer that this cost of communication is proportional to the entropy of node properties (research background). Similarly, interpretations can be derived for other complex systems and can possibly lead to better understanding of the self-organizing principles in the system. 

For specific real-world systems, we can now raise an important question: what optimization process occurs that is beneficial to the functioning/ stability of both the constituents as well as the entire system? Answering this question can help deduce emergence in such systems and perhaps help design a complex system with optimal control on an emerging pattern. Another interesting challenge would be to define the `status' of the constituent entities and relate the entropy of these statuses to the organizational scales and emergent structure of the system.



\section*{Materials and Methods}
 \subsubsection*{Data source}

We analyze multiple real-world networks, datasets for which are openly available: (i) protein structure graph of a protein complex (PSG) \cite{chakrabarty2016naps}, (ii) protein-protein interactions in bacterium \textit{Escherichia coli} (PPI) \cite{kim2015ecolinet}, (iii) gene interaction network of the worm \textit{Caenorhabditis elegans} (GIN) \cite{cho2014wormnet,nr}, (iv) network of nerve fibre tracts in mouse (NER) \cite{bigbrain,nr}, (v) co-authorship network of scientists working on network science (CA) \cite{newman2006finding}, (vi) animal interaction network of weavers (WEA) \cite{nr}, (vii) social network of mutually liked Facebook pages (FB) \cite{rozemberczki2019gemsec,nr} and (viii) infrastructure network of roads connecting cities located mostly in Europe (INF) \cite{bader2012graph,nr}. In the Supplementary material S1, we also analyze the networks of high-throughput protein-protein interactions in \textit{Escherichia coli} (ECO) \cite{kim2015ecolinet} (note that this network is different from the PPI network discussed in the main text in terms of its method of construction and the kind of proteins being considered as nodes) and a network of interactions between drugs (DDI) \cite{biosnapnets}. We discuss that some systems, such as ECO and DDI that exhibit a network structure with limited organizational scales of communities, cannot be reliably characterized by the scaling laws of topological self-similarity discussed in this work.

\subsubsection*{Binary tree representation of hierarchical communities in a network}

\begin{figure}
    \centering
    \includegraphics[width=0.8\textwidth]{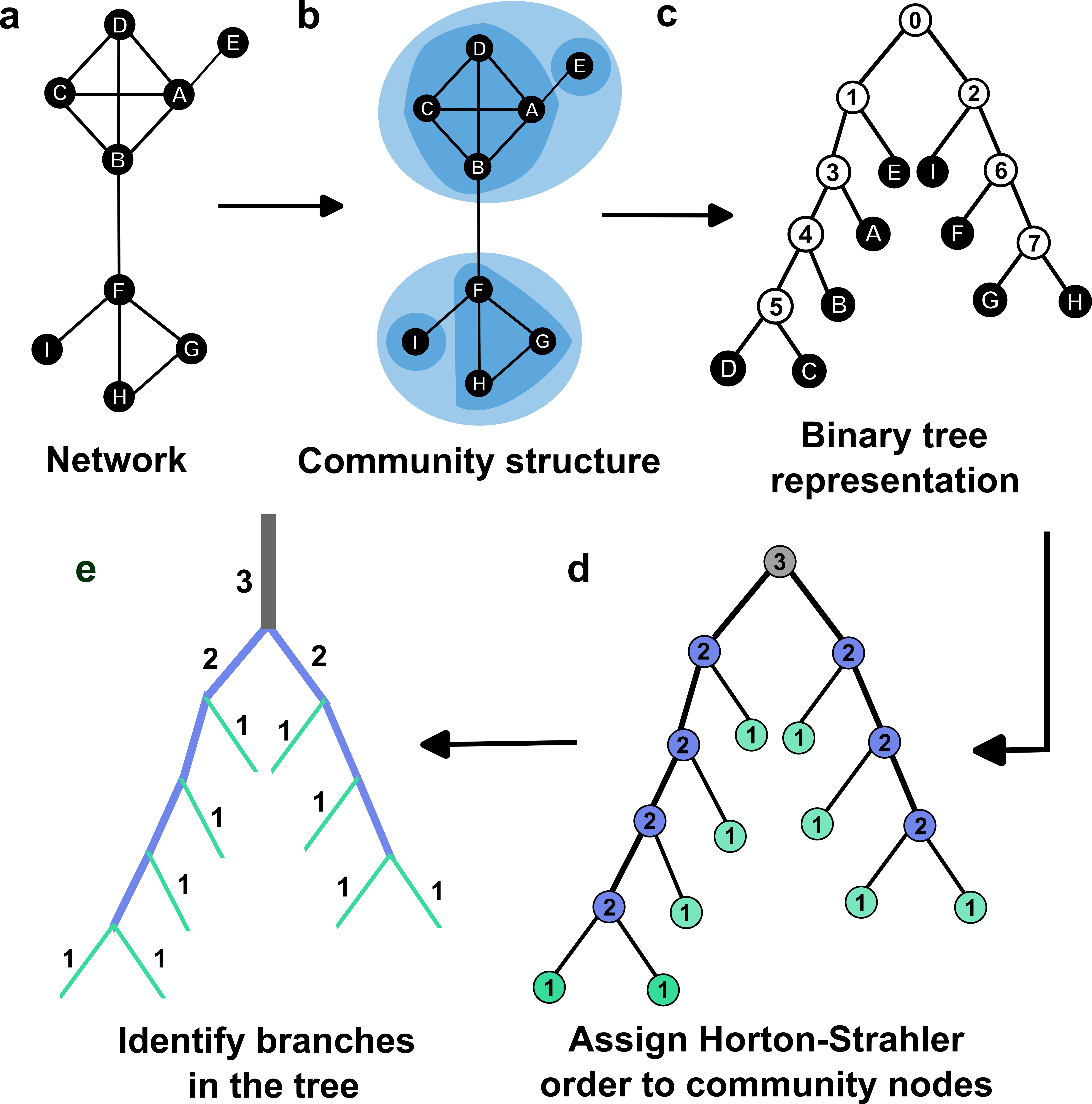}
    \caption{\textbf{Schematic flow diagram for characterizing the hierarchical community structure as a binary tree using the Horton-Strahler indexing scheme.} \textbf{(a)} A sample network. \textbf{(b)} Identifying communities at different scales. \textbf{(c)} The binary tree representation of the structure of the network. Here, community-nodes ($C_j$, labeled as $j$ in the diagram) in the tree represent communities detected at different organizational scales in the complex network shown in (b). Here node `$C_0$' in the tree represents the entire network shown in (a). Two children nodes $C_1$ and $C_2$ are connected to $C_0$ representing the two largest scale communities ($C_1=\text{nodes}\{A,B,C,D,E\},~ C_2=\text{nodes}\{F,G,H,I\}$) evident in (b). Subsequent branching encodes the smaller scale communities within the larger ones. The black community-nodes in the tree correspond to the individual nodes in the network. \textbf{(d)} Horton-Strahler indices assigned to the community-nodes in the tree according to equation \ref{eq_hs}. \textbf{(e)} Identifying branches of different orders in the binary tree representation. The thickness of the lines indicate the Horton-Strahler order $h$. Notice that several lines of same index $h=2$ form a continuous `branch' in the tree. Here, one branch of $h=3$, two branches of $h=2$ and nine branches of $h=1$ exist.}
    \label{fig_flowchart}
\end{figure}

We consider the largest connected component of the undirected, unweighted network representation of a real-world complex system for our analysis. Using Girvan-Newman's algorithm \cite{girvan2002community}, we detect communities at different organizational scales in the network. The algorithm identifies the prominent links (edges) that act as bridges between tightly-knit groups and removes these links to reveal the communities. The edge-betweenness score of a link quantifies the number of times that link falls in the shortest paths between each pair of nodes in the network \cite{freeman1977set}. The link with the highest edge-betweenness score is removed from the network iteratively until two separate disconnected components are obtained. If we use the same algorithm repeatedly, the two large communities detected at first, will split further into smaller and smaller communities. We repeat this process till all the links in the network are removed and the individual nodes of the network separate out as the smallest communities. Through this process, a hierarchy of communities is revealed (see for example figure \ref{fig_flowchart}(a,b)). 

Next, we map the hierarchical communities onto a binary tree representation (see figure \ref{fig_flowchart}(c)). The tree comprises vertices arranged and connected in a hierarchical manner. A vertex is referred as a `community-node' of the tree and represents a community in the complex network. The top node $C_0$ in the tree in figure \ref{fig_flowchart}(c) represents the entire network in figure \ref{fig_flowchart}(b). The two children nodes $C_1$ and $C_2$ descending from $C_0$ in the tree represent the two communities detected in the network at the largest organizational scale. The sub-grouping of these communities at smaller scales is represented by the subsequent branching in the tree. The branching continues until the individual nodes of the network in figure \ref{fig_flowchart}(b) separate out as individual communities represented by black nodes in the tree in figure \ref{fig_flowchart}(c). The tree representation based on community-nodes facilitates an immediate visualization of self-similar branching structure in real-world networks \cite{guimera2003self,arenas2004community}; see Supplementary material S1 for the visualization of self-similar tree representations of multiple real-world networks.

\subsubsection*{Horton-Strahler indexing scheme}

In a network with hierarchical communities, there are many `scales' of organization and each community can be assigned a `hierarchical depth'. We define the hierarchical depth $d_{C_j}$ of a community $C_j$ as the shortest path length between the given community-node $C_j$ and the community-node representing the entire network $C_0$ in the binary tree representation (clearly, $d_{C_0}=0$). To quantify the organizational scales, we use the Horton-Strahler indexing scheme introduced originally for river nets \cite{horton1945erosional, strahler1952dynamic, scheidegger1968horton}. The tree representation of the topology of nested communities in a complex network can be considered analogous to the branching structure of larger rivers into smaller tributaries. 

The community-nodes at the smallest scales, i.e., communities comprising individual nodes of the complex network (colored black in the tree in figure \ref{fig_flowchart}(c)), are assigned $h = 1$. This is the smallest topological scale of the network. Now, consider a community node $C_j$ with two children community-nodes $C_{j,1}$ and $C_{j,2}$ with indices $h_{1}$ and $h_{2}$. Then, the index $h$ for community-node $C_j$ is given by equation \ref{eq_hs}.  
\begin{equation}\label{eq_hs}
            h = \begin{cases} h_1 + 1 & \text { if } h_1 = h_2 \\ 
    \text{max}(h_1,h_2) & \text { otherwise }\end{cases}
        \end{equation}
For the sample network in figure \ref{fig_flowchart}(a), there are three organizational scales ($h=1,2,3$) evident in figure \ref{fig_flowchart}(d). A river analogue of the tree in figure \ref{fig_flowchart}(d) is shown in figure \ref{fig_flowchart}(e), where the thickness of the branch decreases with the organizational scale $h$. The river analogue reveals one thick branch of $h=3$, two branches of $h=2$ and nine branches of $h=1$. Note that, a branch with index $h$ in the tree represents the organizational scale and can run across several levels of the tree. Thus, several communities with similar organizational scale can occur at different hierarchical depths.

\noindent \textbf{Supplementary Material} includes sections S1 - S5.

\noindent \textbf{Acknowledgements:} We are grateful to Dr. Gaurav Chopra and Mr. Ankit Sahay for fruitful discussions.\\
\noindent \textbf{Funding:} This work is funded from J. C. Bose Fellowship (No. JCB/2018/000034/SSC) and the IoE initiative (SP/22-23/1222/CPETWOCTSH). S.T. acknowledges the support from Prime Minister Research Fellowship, Govt. of India. \\
\noindent \textbf{Author Contributions:} S.T.: writing (lead), conception (lead), analysis (equal); N.D.S.: writing (equal), analysis (lead), data handling, plots and network visualization (lead); T.B.: conception (equal), editing (equal), analysis (equal); N.M.: conception (equal), editing (equal); J.K.: conception (equal), editing (equal); R.I.S.: conception (equal), editing (lead), funding (lead). \\
\noindent \textbf{Competing Interests} The authors declare that they have no conflict of interests.\\
\noindent \textbf{Data and materials availability:} Repositories of the data used for analysis are cited in Materials and Methods. All other information necessary to evaluate the conclusions of the paper are described in the paper and the Supplementary Material. Open source codes for constructing a binary tree representation, assigning Horton Strahler index and checking for universal scaling laws are available on DOI: 10.5281/zenodo.15879191. Codes for generating a network through status model described in this work are available on DOI: 10.5281/zenodo.15879122.




\newpage
 
\renewcommand{\thesection}{S\arabic{section}}
\section*{Supplementary Material}
\beginsupplement
\def\smgb{$-0.51 (1.00)$}
\def\smgx{$-0.45 (0.98)$}
\def\smge{$-0.42 (1.00)$}
\def\smgn{$0.50 (1.00)$}
\def\smgh{$-0.07 (0.98)$}
\def\smgxd{$-0.38 (1.00)$}
\def\smged{$-0.42 (0.98)$}
\def\smgnd{$0.38 (1.00)$}
\def\realgb{$-0.53\pm0.04$}
\def\realgx{$-0.46\pm0.03$}
\def\realge{$-0.44\pm0.03$}
\def\realgn{$0.50\pm0.03$}
\def\realgh{$-0.08\pm0.01$}
\def\realgxd{$-0.37\pm0.03$}
\def\realged{$-0.39\pm0.04$}
\def\realgnd{$0.37\pm0.03$}
\def\psggb{$-0.44 (0.97)$}
\def\psggx{$-0.49 (0.98)$}
\def\psgge{$-0.37 (1.00)$}
\def\psggn{$0.45 (0.99)$}
\def\psggh{$-0.08 (0.99)$}
\def\psggxd{$-0.37 (0.99)$}
\def\psgged{$-0.35 (0.99)$}
\def\psggnd{$0.37 (0.99)$}
\def\ppigb{$-0.55 (1.00)$}
\def\ppigx{$-0.45 (0.88)$}
\def\ppige{$-0.48 (1.00)$}
\def\ppign{$0.48 (0.96)$}
\def\ppigh{$-0.09 (0.98)$}
\def\ppigxd{$-0.38 (0.99)$}
\def\ppiged{$-0.43 (0.90)$}
\def\ppignd{$0.38 (0.99)$}
\def\gingb{$-0.59 (0.99)$}
\def\gingx{$-0.40 (0.90)$}
\def\ginge{$-0.48 (1.00)$}
\def\gingn{$0.57 (0.97)$}
\def\gingh{$-0.08 (0.89)$}
\def\gingxd{$-0.29 (0.91)$}
\def\ginged{$-0.38 (0.78)$}
\def\gingnd{$0.29 (0.91)$}
\def\nergb{$-0.61 (0.99)$}
\def\nergx{$-0.41 (0.77)$}
\def\nerge{$-0.48 (1.00)$}
\def\nergn{$0.54 (0.99)$}
\def\nergh{$-0.09 (0.95)$}
\def\nergxd{$-0.31 (0.98)$}
\def\nerged{$-0.40 (0.88)$}
\def\nergnd{$0.31 (0.98)$}
\def\cagb{$-0.52 (0.99)$}
\def\cagx{$-0.51 (0.96)$}
\def\cage{$-0.46 (0.99)$}
\def\cagn{$0.50 (1.00)$}
\def\cagh{$-0.08 (0.99)$}
\def\cagxd{$-0.42 (0.99)$}
\def\caged{$-0.42 (0.94)$}
\def\cagnd{$0.42 (0.99)$}
\def\weagb{$-0.50 (0.99)$}
\def\weagx{$-0.48 (0.97)$}
\def\weage{$-0.38 (1.00)$}
\def\weagn{$0.48 (0.98)$}
\def\weagh{$-0.10 (0.98)$}
\def\weagxd{$-0.40 (1.00)$}
\def\weaged{$-0.25 (0.77)$}
\def\weagnd{$0.41 (1.00)$}
\def\fbgb{$-0.56 (0.98)$}
\def\fbgx{$-0.52 (0.91)$}
\def\fbge{$-0.47 (0.99)$}
\def\fbgn{$0.52 (0.92)$}
\def\fbgh{$-0.09 (0.99)$}
\def\fbgxd{$-0.39 (1.00)$}
\def\fbged{$-0.38 (0.91)$}
\def\fbgnd{$0.39 (1.00)$}
\def\infgb{$-0.44 (0.99)$}
\def\infgx{$-0.39 (0.99)$}
\def\infge{$-0.41 (0.99)$}
\def\infgn{$0.42 (1.00)$}
\def\infgh{$-0.06 (0.99)$}
\def\infgxd{$-0.42 (0.99)$}
\def\infged{$-0.49 (0.99)$}
\def\infgnd{$0.42 (0.99)$}
\def\ecogb{$-0.83(0.94)$}
\def\ecogx{$-0.32(0.96)$}
\def\ecoge{$-0.72(0.98)$}
\def\ecogn{$0.8(0.93)$}
\def\ecogh{$-0.08(0.97)$}
\def\ecogxd{$-0.39(0.99)$}
\def\ecoged{$-0.53(0.69)$}
\def\ecognd{$0.39(0.99)$}
\def\ddigb{$-1.03(0.92)$}
\def\ddigx{$-0.22(0.98)$}
\def\ddige{$-0.54(0.99)$}
\def\ddign{$1.03(0.89)$}
\def\ddigh{$-0.06(0.84)$}
\def\ddigxd{$-0.43(0.96)$}
\def\ddiged{$-0.37(0.74)$}
\def\ddignd{$0.43(0.96)$}
\def\statgb{$-0.48\pm0.04$}
\def\statgx{$-0.43\pm0.02$}
\def\statge{$-0.38\pm0.03$}
\def\statgn{$0.47\pm0.03$}
\def\statgh{$-0.08\pm0.00$}
\def\statgxd{$-0.36\pm0.02$}
\def\statged{$-0.39\pm0.02$}
\def\statgnd{$0.36\pm0.02$}
\def\qgb{$-0.44(1.00)$}
\def\qgx{$-0.43(0.98)$}
\def\qge{$-0.36(1.00)$}
\def\qgn{$0.43(1.00)$}
\def\qgh{$-0.07(0.99)$}
\def\qgxd{$-0.36(1.00)$}
\def\qged{$-0.37(0.99)$}
\def\qgnd{$0.36(1.00)$}
\def\egb{$-0.5(1.00)$}
\def\egx{$-0.42(0.99)$}
\def\ege{$-0.38(1.00)$}
\def\egn{$0.47(1.00)$}
\def\egh{$-0.08(0.98)$}
\def\egxd{$-0.35(0.99)$}
\def\eged{$-0.39(0.96)$}
\def\egnd{$0.35(0.99)$}
\def\ggb{$-0.51(1.00)$}
\def\ggx{$-0.45(0.98)$}
\def\gge{$-0.42(1.00)$}
\def\ggn{$0.5(1.00)$}
\def\ggh{$-0.07(0.98)$}
\def\ggxd{$-0.38(1.00)$}
\def\gged{$-0.42(0.98)$}
\def\ggnd{$0.38(1.00)$}
\def\ngb{$-0.53\pm0.08$}
\def\ngx{$-0.49\pm0.04$}
\def\nge{$-0.4\pm0.04$}
\def\ngn{$0.52\pm0.07$}
\def\ngh{$-0.08\pm0.01$}
\def\ngxd{$-0.32\pm0.04$}
\def\nged{$-0.38\pm0.04$}
\def\ngnd{$0.33\pm0.04$}
\def\nagb{$-0.51(1.00)$}
\def\nagx{$-0.45(0.98)$}
\def\nage{$-0.42(1.00)$}
\def\nagn{$0.5(1.00)$}
\def\nagh{$-0.07(0.98)$}
\def\nagxd{$-0.38(1.00)$}
\def\naged{$-0.42(0.98)$}
\def\nagnd{$0.38(1.00)$}
\def\nbgb{$-0.57(0.99)$}
\def\nbgx{$-0.52(0.94)$}
\def\nbge{$-0.43(1.00)$}
\def\nbgn{$0.55(1.00)$}
\def\nbgh{$-0.08(0.99)$}
\def\nbgxd{$-0.37(0.99)$}
\def\nbged{$-0.42(0.94)$}
\def\nbgnd{$0.37(0.99)$}
\def\ncgb{$-0.58(0.99)$}
\def\ncgx{$-0.51(0.88)$}
\def\ncge{$-0.42(0.99)$}
\def\ncgn{$0.57(0.99)$}
\def\ncgh{$-0.09(0.97)$}
\def\ncgxd{$-0.29(0.99)$}
\def\ncged{$-0.37(0.89)$}
\def\ncgnd{$0.29(0.99)$}
\def\npsgb{$-0.58(0.99)$}
\def\npsgx{$-0.38(0.96)$}
\def\npsge{$-0.43(0.99)$}
\def\npsgn{$0.54(0.99)$}
\def\npsgh{$-0.09(0.99)$}
\def\npsgxd{$-0.37(1.00)$}
\def\npsged{$-0.38(0.85)$}
\def\npsgnd{$0.37(1.00)$}
\def\nmsgb{$-0.41(1.00)$}
\def\nmsgx{$-0.36(1.00)$}
\def\nmsge{$-0.39(1.00)$}
\def\nmsgn{$0.39(1.00)$}
\def\nmsgh{$-0.06(0.99)$}
\def\nmsgxd{$-0.38(0.98)$}
\def\nmsged{$-0.43(0.95)$}
\def\nmsgnd{$0.38(0.98)$}
\def\mgb{$-0.49\pm0.04$}
\def\mgx{$-0.43\pm0.03$}
\def\mge{$-0.37\pm0.01$}
\def\mgn{$0.48\pm0.04$}
\def\mgh{$-0.07\pm0.02$}
\def\mgxd{$-0.28\pm0.03$}
\def\mged{$-0.31\pm0.07$}
\def\mgnd{$0.28\pm0.03$}
\def\magb{$-0.44(1.00)$}
\def\magx{$-0.37(0.99)$}
\def\mage{$-0.39(1.00)$}
\def\magn{$0.42(1.00)$}
\def\magh{$-0.08(0.99)$}
\def\magxd{$-0.33(0.98)$}
\def\maged{$-0.39(0.94)$}
\def\magnd{$0.33(0.98)$}
\def\mbgb{$-0.51(1.00)$}
\def\mbgx{$-0.45(0.98)$}
\def\mbge{$-0.42(1.00)$}
\def\mbgn{$0.5(1.00)$}
\def\mbgh{$-0.07(0.98)$}
\def\mbgxd{$-0.38(1.00)$}
\def\mbged{$-0.42(0.98)$}
\def\mbgnd{$0.38(1.00)$}
\def\mcgb{$-0.51(0.99)$}
\def\mcgx{$-0.45(0.98)$}
\def\mcge{$-0.39(1.00)$}
\def\mcgn{$0.51(1.00)$}
\def\mcgh{$-0.02(0.24)$}
\def\mcgxd{$-0.22(0.42)$}
\def\mcged{$-0.15(0.07)$}
\def\mcgnd{$0.22(0.42)$}
\def\mdgb{$-0.5(0.99)$}
\def\mdgx{$-0.47(0.95)$}
\def\mdge{$-0.36(1.00)$}
\def\mdgn{$0.5(0.99)$}
\def\mdgh{$-0.07(0.88)$}
\def\mdgxd{$-0.26(0.89)$}
\def\mdged{$-0.34(0.97)$}
\def\mdgnd{$0.26(0.89)$}
\def\megb{$-0.58(0.99)$}
\def\megx{$-0.44(0.98)$}
\def\mege{$-0.37(0.99)$}
\def\megn{$0.54(0.99)$}
\def\megh{$-0.07(0.95)$}
\def\megxd{$-0.3(0.96)$}
\def\meged{$-0.32(0.94)$}
\def\megnd{$0.3(0.96)$}

Complexity in real-world systems is intriguing for two reasons: (i) universality in the emergent topology of the systems, and (ii) variability across systems despite such universality. Here, we present a detailed analysis of multiple real-world networks and variations of the phenomenological model presented in the main text.

We present the degree distributions and adjacency matrices for multiple real-world networks that exhibit topological self-similarity (Sec. \ref{suppsec_real_world}). We also examine examples of networks which do not exhibit such self-similar topology of communities. Next, (Sec. \ref{suppsec_vary_status}) we investigate the network structure and self-similarity of hierarchical communities when the underlying distribution of statuses of nodes is changed in the model. We show that for any non-uniform status distribution, the resulting network exhibits self-similar hierarchical communities. Moreover, we show that the process of self-organization entails the same optimization process as discussed in the main text irrespective of the underlying status distribution. Finally, in Sec. \ref{suppsec_vary_m} and Sec. \ref{suppsec_growth} we discuss the effect of link density (or parameter $m$) and growth (parameter $N$) on the hierarchical organization in networks obtained from the model. We also present an alternative non-growing version of the model and demonstrate that self-similar hierarchical communities can be formed in non-growing network models.

\section{Self-similarity of hierarchical community structure in real-world networks} \label{suppsec_real_world}

We have considered ten real-world networks from different domains such as social, biological, infrastructural systems, etc. In particular, we study the structure of (a) a protein complex (PSG), (b) protein-protein interactions in bacterium \textit{Escherichia coli} (PPI), (c) gene interaction network of the worm \textit{Caenorhabditis elegans} (GIN), (d) network of nerve fibre tracts in mouse (NER), (e) coauthor network of researchers in network science (CA), (f) animal interaction network of weavers (WEA), (g) social network of mutually liked Facebook pages (FB), (h) infrastructure network of roads connecting cities in mostly Europe (INF), (i) high-throughput protein-protein interactions in \textit{Escherichia coli} (ECO), and (j) drug-drug interactions (DDI). 

\begin{table}[]
\centering
\resizebox{\textwidth}{!}{%
\begin{tabular}{|c|c|rrrrrrrrrr|}
\hline
\multirow{2}{*}{Coefficient} & \multirow{2}{*}{$y$ versus $x$}                                & \multicolumn{10}{c|}{Real-world networks}                                                                                                                                                                                                                                                                                                                                                                                                            \\ \cline{3-12} 
                             &                                                                & \multicolumn{1}{c|}{PSG}                    & \multicolumn{1}{c|}{PPI}                    & \multicolumn{1}{c|}{GIN}                    & \multicolumn{1}{c|}{NER}                    & \multicolumn{1}{c|}{CA}                    & \multicolumn{1}{c|}{WEA}                    & \multicolumn{1}{c|}{FB}                    & \multicolumn{1}{c|}{INF}                    & \multicolumn{1}{c|}{ECO}                    & \multicolumn{1}{c|}{DDI} \\ \hline
$\gamma_b$                   & $\log_{10}(b_h)$ vs $h$                                        & \multicolumn{1}{r|}{\psggb}  & \multicolumn{1}{r|}{\ppigb}  & \multicolumn{1}{r|}{\gingb}  & \multicolumn{1}{r|}{\nergb}  & \multicolumn{1}{r|}{\cagb}  & \multicolumn{1}{r|}{\weagb}  & \multicolumn{1}{r|}{\fbgb}  & \multicolumn{1}{r|}{\infgb}  & \multicolumn{1}{r|}{\ecogb}  & \ddigb    \\ \hline
$\gamma_\chi$                & $\log_{10}(\chi_h)$ vs $h$                                     & \multicolumn{1}{r|}{\psggx}  & \multicolumn{1}{r|}{\ppigx}  & \multicolumn{1}{r|}{\gingx}  & \multicolumn{1}{r|}{\nergx}  & \multicolumn{1}{r|}{\cagx}  & \multicolumn{1}{r|}{\weagx}  & \multicolumn{1}{r|}{\fbgx}  & \multicolumn{1}{r|}{\infgx}  & \multicolumn{1}{r|}{\ecogx}  & \ddigx    \\ \hline
$\gamma_\eta$                & $\log_{10}(\langle \eta \rangle _h)$ vs $h$                    & \multicolumn{1}{r|}{\psgge}  & \multicolumn{1}{r|}{\ppige}  & \multicolumn{1}{r|}{\ginge}  & \multicolumn{1}{r|}{\nerge}  & \multicolumn{1}{r|}{\cage}  & \multicolumn{1}{r|}{\weage}  & \multicolumn{1}{r|}{\fbge}  & \multicolumn{1}{r|}{\infge}  & \multicolumn{1}{r|}{\ecoge}  & \ddige    \\ \hline
$\gamma_n$                   & $\log_{10}(\langle n \rangle _h)$ vs $h$                       & \multicolumn{1}{r|}{\psggn}  & \multicolumn{1}{r|}{\ppign}  & \multicolumn{1}{r|}{\gingn}  & \multicolumn{1}{r|}{\nergn}  & \multicolumn{1}{r|}{\cagn}  & \multicolumn{1}{r|}{\weagn}  & \multicolumn{1}{r|}{\fbgn}  & \multicolumn{1}{r|}{\infgn}  & \multicolumn{1}{r|}{\ecogn}  & \ddign    \\ \hline
$\gamma_{h}$                 & $\log_{10}(\langle h \rangle _d)$ vs $d$                       & \multicolumn{1}{r|}{\psggh}  & \multicolumn{1}{r|}{\ppigh}  & \multicolumn{1}{r|}{\gingh}  & \multicolumn{1}{r|}{\nergh}  & \multicolumn{1}{r|}{\cagh}  & \multicolumn{1}{r|}{\weagh}  & \multicolumn{1}{r|}{\fbgh}  & \multicolumn{1}{r|}{\infgh}  & \multicolumn{1}{r|}{\ecogh}  & \ddigh    \\ \hline
$\gamma_{\chi_d}$            & $\log_{10}(\chi _d)$ vs $\langle h \rangle _d$                 & \multicolumn{1}{r|}{\psggxd} & \multicolumn{1}{r|}{\ppigxd} & \multicolumn{1}{r|}{\gingxd} & \multicolumn{1}{r|}{\nergxd} & \multicolumn{1}{r|}{\cagxd} & \multicolumn{1}{r|}{\weagxd} & \multicolumn{1}{r|}{\fbgxd} & \multicolumn{1}{r|}{\infgxd} & \multicolumn{1}{r|}{\ecogxd} & \ddigxd   \\ \hline
$\gamma_{\eta_d}$            & $\log_{10}(\langle \eta \rangle _d)$ vs $\langle h \rangle _d$ & \multicolumn{1}{r|}{\psgged} & \multicolumn{1}{r|}{\ppiged} & \multicolumn{1}{r|}{\ginged} & \multicolumn{1}{r|}{\nerged} & \multicolumn{1}{r|}{\caged} & \multicolumn{1}{r|}{\weaged} & \multicolumn{1}{r|}{\fbged} & \multicolumn{1}{r|}{\infged} & \multicolumn{1}{r|}{\ecoged} & \ddiged   \\ \hline
$\gamma_{n_d}$               & $\log_{10}(\langle n \rangle _d)$ vs $\langle h \rangle _d$    & \multicolumn{1}{r|}{\psggnd} & \multicolumn{1}{r|}{\ppignd} & \multicolumn{1}{r|}{\gingnd} & \multicolumn{1}{r|}{\nergnd} & \multicolumn{1}{r|}{\cagnd} & \multicolumn{1}{r|}{\weagnd} & \multicolumn{1}{r|}{\fbgnd} & \multicolumn{1}{r|}{\infgnd} & \multicolumn{1}{r|}{\ecognd} & \ddignd   \\ \hline
\end{tabular}%
}
\caption{Horton scaling exponents observed for different real-world networks including the ones which do not exhibit topological self-similarity (ECO and DDI networks). The scaling exponents are listed with the goodness of fit (R-Square) in brackets.}
\label{table_supp_realworld}
\end{table}

\begin{figure}
    \centering
    \includegraphics[width=\textwidth]{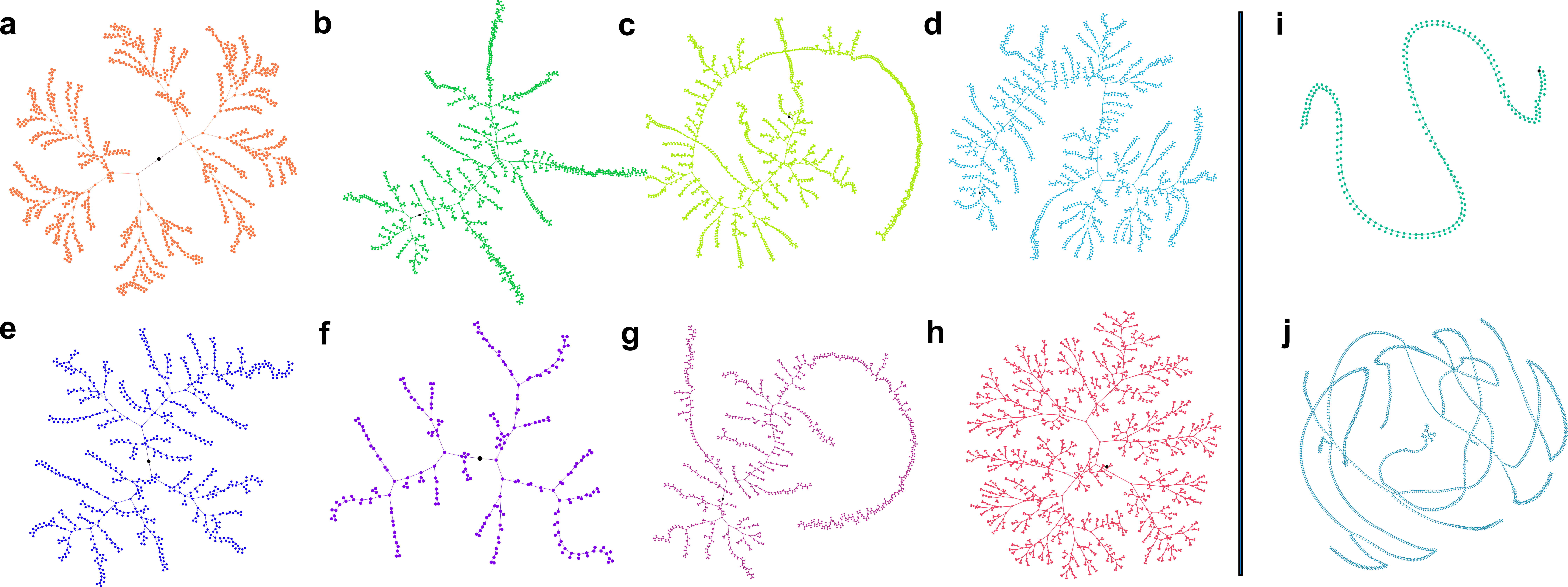}
    \caption{\textbf{Binary trees representing the hierarchical organization of communities of different real-world networks}: (a) protein structure graph of a protein complex (PSG), (b) protein-protein interactions in bacterium \textit{Escherichia coli} (PPI), (c) gene interaction network of the worm \textit{Caenorhabditis elegans} (GIN), (d) network of nerve fibre tracts in mouse (NER), (e) coauthorship network of scientists working on network science (CA), (f) interaction network of weavers (WEA), (g) network of mutually liked Facebook pages (FB), (h) infrastructure network of roads connecting cities in mostly Europe (INF), (i) high-throughput protein-protein interactions in \textit{Escherichia coli} (ECO), and (j) drug-drug interactions (DDI). The black node in the tree corresponds to the entire network. Notice, the trees in (a) to (h) delineate self-similar branching structure whereas such branching structure is not evident in (i,j). }
    \label{fig_tree_all}
\end{figure}

We visualize the binary tree representation of these networks in Fig. \ref{fig_tree_all}. Clearly, the network examples in Fig. \ref{fig_tree_all}(a-h) exhibit self-similar branching structure. Table \ref{table_supp_realworld} shows that examples (a) to (h) exhibit the universal scaling relations describing the self-similar hierarchical community structure. In other words, these networks follow Horton's law and also display structural self-similarity through scaling relations obtained at fixed hierarchical depths. The scaling exponents thus obtained are universal, i.e., similar across all the examples in Fig. \ref{fig_tree_all}(a-h).

However, the tree representations of the ECO and DDI networks in Fig. \ref{fig_tree_all}(i,j) are not structurally self-similar. Groups of very few nodes or individual nodes separate out from the network as communities represented by short branches along a long chain-like structure in the tree representation. Thus, the ECO and DDI networks exhibit very few organizational scales. These networks do exhibit the Horton's law of branch numbers and mean attributes; however, the scaling exponents are very different from that for the universal scaling relations reported for networks in Fig. \ref{fig_tree_all}(a-h). Moreover, scaling exponent $\gamma_{\eta_d}$ quantifying the variation of the link density of communities at fixed hierarchical depth with $\langle h \rangle _d$ has a poor goodness-of-fit. Clearly, the ECO and DDI networks are Hortonian but not structurally self-similar and hence do not obey the universal scaling relations.

\begin{figure}
    \centering
    \includegraphics[width=\textwidth]{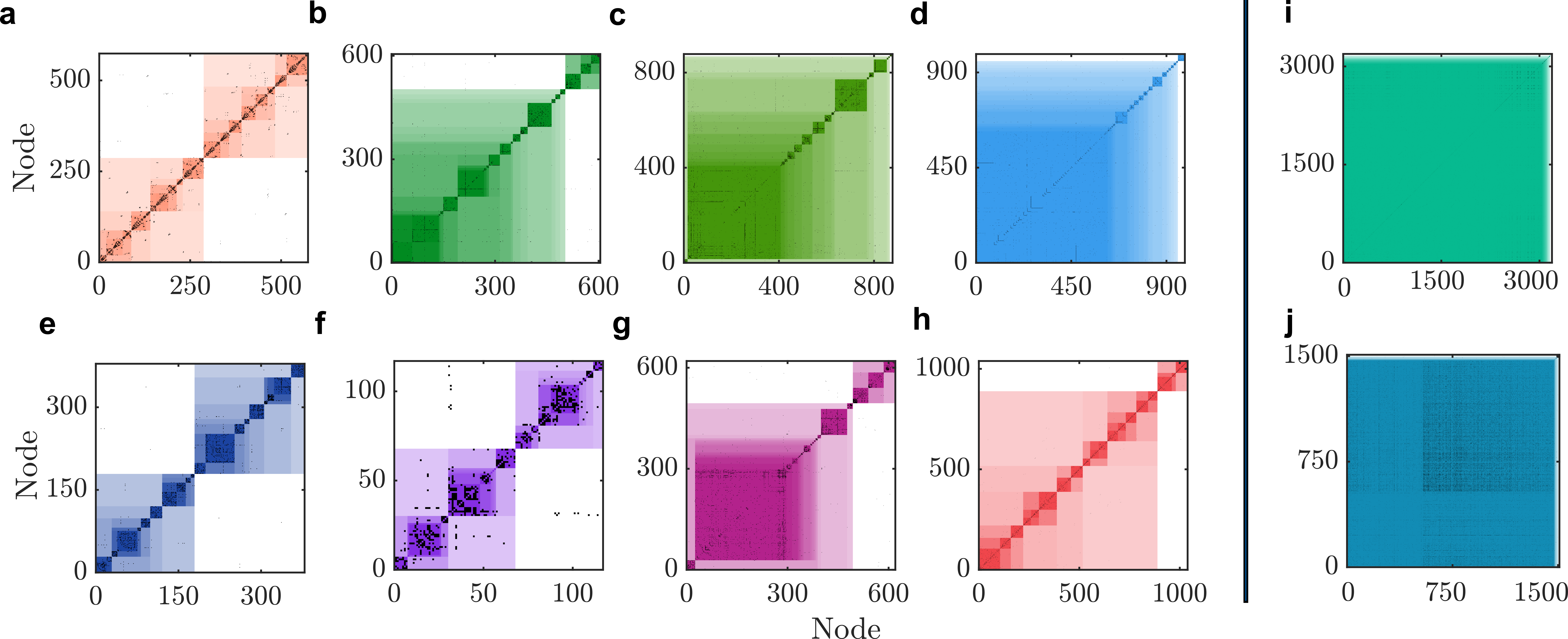}
    \caption{\textbf{Adjacency matrices showing the community-within-community structure of different real-world networks:} (a) protein structure graph of a protein complex (PSG), (b) protein-protein interactions in bacterium \textit{Escherichia coli} (PPI), (c) gene interaction network of the worm \textit{Caenorhabditis elegans} (GIN), (d) network of nerve fibre tracts in mouse (NER), (e) coauthorship network of scientists working on network science (CA), (f) interaction network of weavers (WEA), (g) network of mutually liked Facebook pages (FB), (h) infrastructure network of roads connecting cities in mostly Europe (INF), (i) high-throughput protein-protein interactions in \textit{Escherichia coli} (ECO), and (j) drug-drug interactions (DDI). The colored boxes represent the communities obtained at different scales, smaller scales are represented by increasing opacity of the color. Notice, a clear community-within-community structure is evident in (a) to (h), but not in (i,j). Also, the community-within-community structure is distinct for each network in (a) to (h) depicting the variability in topology of hierarchical communities across diverse networks.}
    \label{fig_s1_adj}
\end{figure}

Next, we visualize the adjacency matrices of each of the networks. Such an adjacency matrix is obtained after rearranging the nodes, such that, nodes in the same community are located in vicinity of each other in the matrix. Such rearrangement of nodes is done at multiple scales and helps visualize the community-within-community structure and the composition of communities that split at multiple scales. The adjacency matrices of the networks ECO and DDI in Fig. \ref{fig_s1_adj}(i) and (j) respectively show that these networks have no community structure. The large block structures that span across several scales do not represent communities but are just remnants of the community detection algorithm as individual nodes/very small group of nodes separate out from the network during each iteration.

On the other hand, the adjacency matrices of networks (a-h) exhibit unique block structures repeated at different scales. Notice the difference between the composition of hierarchical communities across examples in Fig. \ref{fig_s1_adj}(a-h). For instance, a large-scale community in Fig. \ref{fig_s1_adj}(a,f) splits into constituent communities of comparable sizes. On the other hand, a large-scale community in Fig. \ref{fig_s1_adj}(b,h) splits into two communities of relatively distinct sizes (approximately a ratio of 80-20 percent nodes). Yet, all of these networks depicted in Fig. \ref{fig_s1_adj}(a-h) exhibit the universal scaling laws of structural self-similarity. Hence, we infer that universal scaling laws obtained from the tree representation imply similar organizational processes across multiple real-world networks. And yet, there is room for case-to-case variability in the composition and distribution of communities across scales. It is fascinating how these diverse networks having differences in the community-within-community composition can obey the same scaling laws owing to universality in the topological structure. 

\begin{figure}
    \centering
    \includegraphics[width=\textwidth]{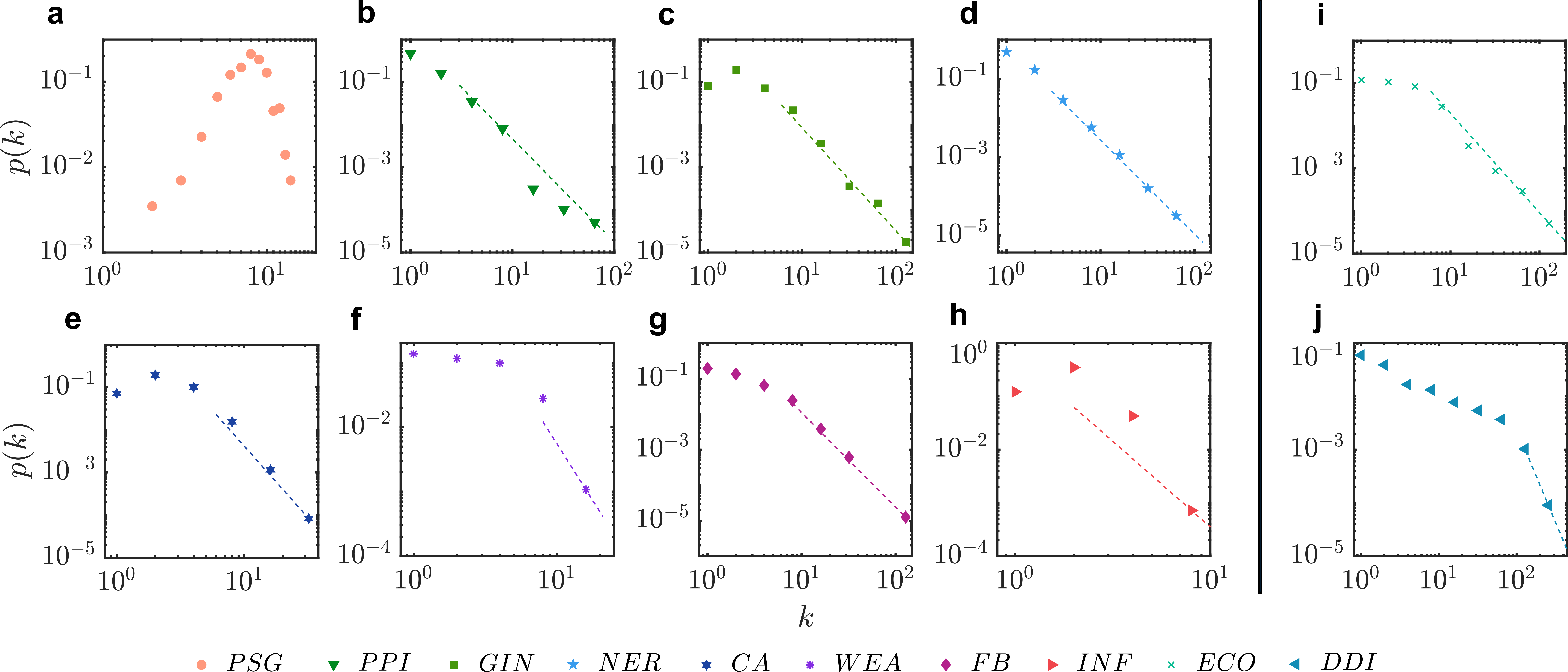}
    \caption{\textbf{Degree distribution of different real-world networks}: (a) protein structure graph of a protein complex (PSG), (b) protein-protein interactions in bacterium \textit{Escherichia coli} (PPI), (c) gene interaction network of the worm \textit{Caenorhabditis elegans} (GIN), (d) network of nerve fibre tracts in mouse (NER), (e) coauthorship network of scientists working on network science (CA), (f) interaction network of weavers (WEA), (g) network of mutually liked Facebook pages (FB), (h) infrastructure network of roads connecting cities in mostly Europe (INF), (i) high-throughput protein-protein interactions in \textit{Escherichia coli} (ECO), and (j) drug-drug interactions (DDI). }
    \label{fig_s1_deg}
\end{figure}

Finally, we examine the degree distribution of the real-world networks that organize into a hierarchical community structure (see Fig. \ref{fig_s1_deg}). The degree distribution of PSG exhibits a peculiar degree distribution that appears to be log-normal in nature (Fig. \ref{fig_s1_deg}(a)). Further, some networks exhibit scale-free degree distribution such as in Fig. \ref{fig_s1_deg}(b,d). Other networks (see Fig. \ref{fig_s1_deg}(c, e-j)) exhibit a power-law like degree distribution where the power law exponent may vary with the range of degree \cite{newman2001scientific, qian2021emergence}. Clearly, despite universality in the emergent topology, the degree distributions of the diverse real-world networks are not similar. Further, similarity in degree distribution does not imply similarity in the emergent topology; specifically, we note that ECO and FB networks exhibit similar degree distribution but are very different in topological structure.

In the next section, using our model we explain the variability in the community decomposition and distribution in networks exhibiting universality in topological structure.

\section{Effect of underlying status distribution on the topology and composition of hierarchical communities}\label{suppsec_vary_status}

\begin{figure}
    \centering
    \includegraphics[width=0.5\textwidth]{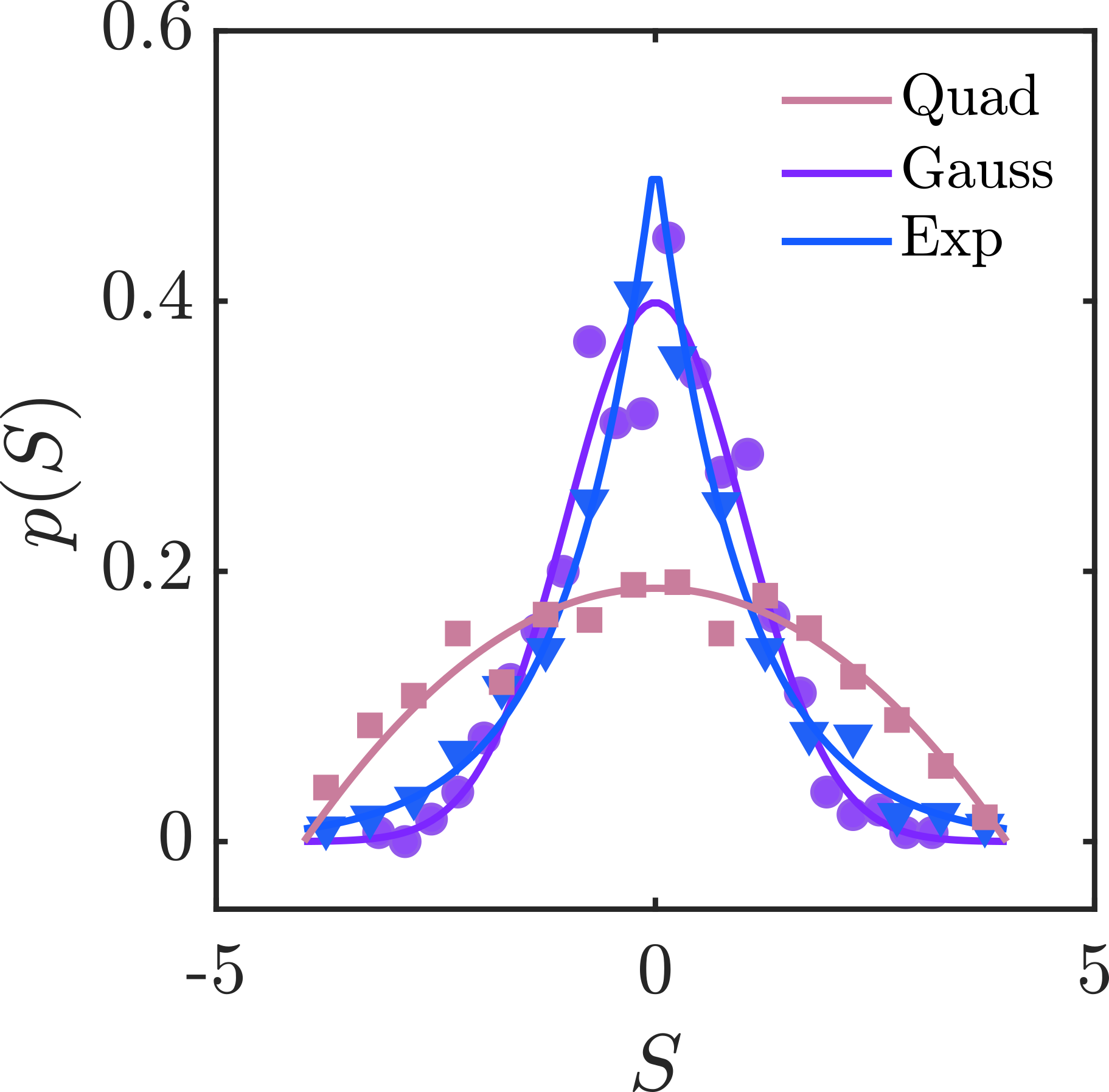}
    \caption{\textbf{Different status distributions used to construct networks from the status model.} Quadratic, exponential and Gaussian probability density functions with same mean ($\mu = 0$) and similar range ($\sim [-4,4]$) of statuses.}
    \label{fig_s2_status}
\end{figure}

Here, we discuss differences and similarity in the hierarchical community structure obtained from our model when the underlying status distributions are distinct. We consider a Gaussian, a quadratic and an exponential distribution as shown in Fig. \ref{fig_s2_status}.

Firstly, we examine the topological structure of the networks obtained from different status distributions. We find that the tree representations of these networks exhibit self-similar branching structure, obey the universal scaling relations and are structurally self-similar across various hierarchical depths (see Table \ref{table_supp_vary_status}, all exponents have high R-square values). Thus, we infer that despite distinct underlying status distributions, the emergent topology of hierarchical communities in networks is universal.


\begin{table}[]
\centering
\begin{tabular}{|c|c|rrrr|}
\hline
\multirow{2}{*}{Coefficient} & \multirow{2}{*}{$y$ versus $x$}                                & \multicolumn{4}{c|}{Status model networks - varying status distribution}                                                            \\ \cline{3-6} 
                             &                                                                & \multicolumn{1}{c|}{Quadratic} & \multicolumn{1}{c|}{Exponential} & \multicolumn{1}{c|}{Gaussian} & \multicolumn{1}{c|}{Universal law} \\ \hline
$\gamma_b$                   & $\log_{10}(b_h)$ vs $h$                                        & \multicolumn{1}{r|}{\qgb}      & \multicolumn{1}{r|}{\egb}        & \multicolumn{1}{r|}{\ggb}     & \statgb                         \\ \hline
$\gamma_\chi$                & $\log_{10}(\chi_h)$ vs $h$                                     & \multicolumn{1}{r|}{\qgx}      & \multicolumn{1}{r|}{\egx}        & \multicolumn{1}{r|}{\ggx}     & \statgx                         \\ \hline
$\gamma_\eta$                & $\log_{10}(\langle \eta \rangle _h)$ vs $h$                    & \multicolumn{1}{r|}{\qge}      & \multicolumn{1}{r|}{\ege}        & \multicolumn{1}{r|}{\gge}     & \statge                         \\ \hline
$\gamma_n$                   & $\log_{10}(\langle n \rangle _h)$ vs $h$                       & \multicolumn{1}{r|}{\qgn}      & \multicolumn{1}{r|}{\egn}        & \multicolumn{1}{r|}{\ggn}     & \statgn                         \\ \hline
$\gamma_{h}$                 & $\log_{10}(\langle h \rangle _d)$ vs $d$                       & \multicolumn{1}{r|}{\qgh}      & \multicolumn{1}{r|}{\egh}        & \multicolumn{1}{r|}{\ggh}     & \statgh                         \\ \hline
$\gamma_{\chi_d}$            & $\log_{10}(\chi _d)$ vs $\langle h \rangle _d$                 & \multicolumn{1}{r|}{\qgxd}     & \multicolumn{1}{r|}{\egxd}       & \multicolumn{1}{r|}{\ggxd}    & \statgxd                        \\ \hline
$\gamma_{\eta_d}$            & $\log_{10}(\langle \eta \rangle _d)$ vs $\langle h \rangle _d$ & \multicolumn{1}{r|}{\qged}     & \multicolumn{1}{r|}{\eged}       & \multicolumn{1}{r|}{\gged}    & \statged                        \\ \hline
$\gamma_{n_d}$               & $\log_{10}(\langle n \rangle _d)$ vs $\langle h \rangle _d$    & \multicolumn{1}{r|}{\qgnd}     & \multicolumn{1}{r|}{\egnd}       & \multicolumn{1}{r|}{\ggnd}    & \statgnd                        \\ \hline
\end{tabular}%
\caption{Horton scaling exponents with goodness of fit (R-square values in brackets) for networks obtained through the status model, each having a different input status distribution. The mean exponents for all these cases are also mentioned in the last column along with the standard error with $90\%$ confidence. In particular, we analyze networks with quadratic, exponential and Gaussian status distribution with the same input parameters $N = 1000$, $m = 4$ and $N_0 = 100$.}
\label{table_supp_vary_status}
\end{table}

\begin{figure}
    \centering
    \includegraphics[width=\textwidth]{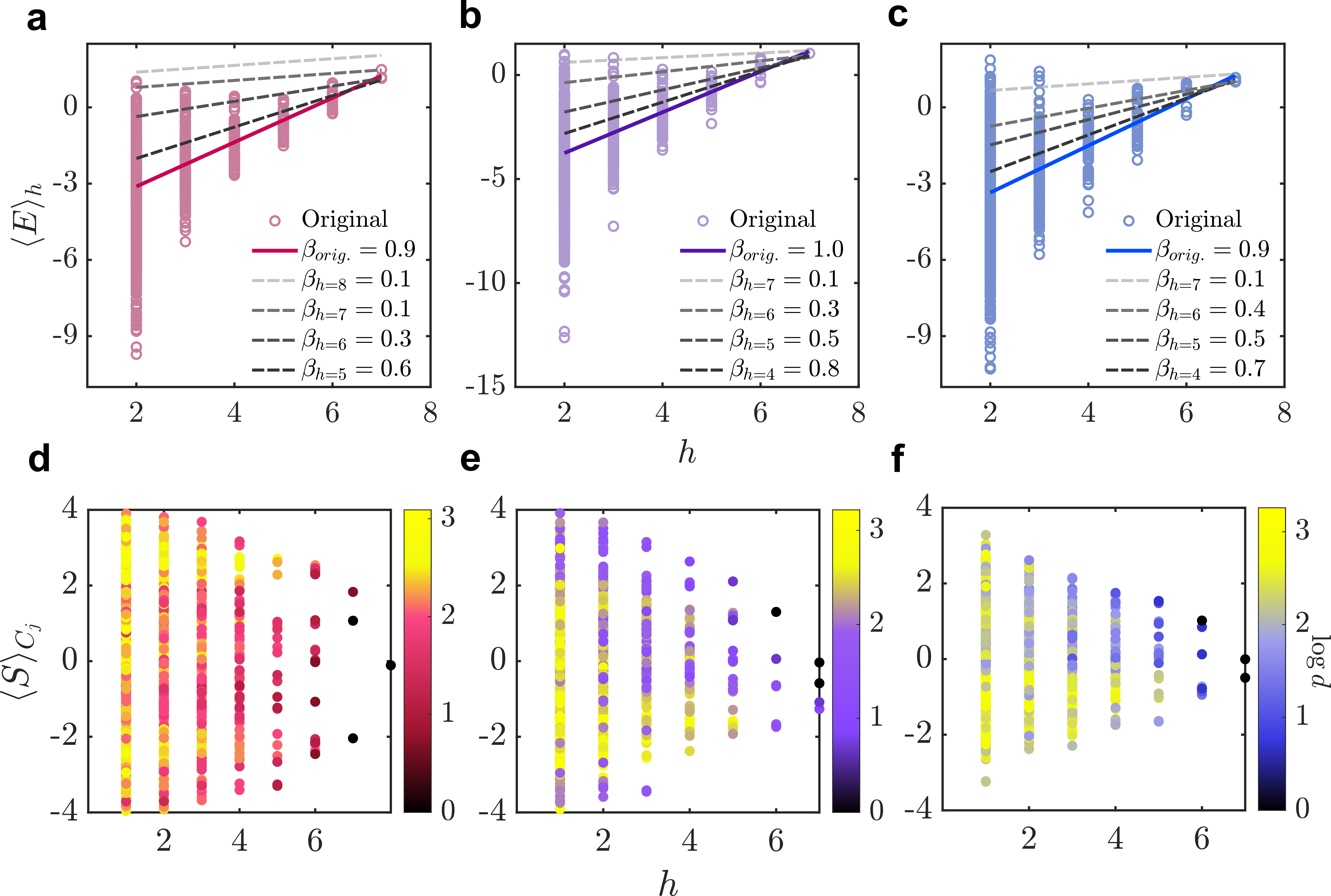}
    \caption{\textbf{The process of self-organization in networks obtained from the status model having different status distributions is universal.} (a-c) Variation of mean status entropy at a constant order $h$ for networks obtained through the status model, each network having a different status distribution. In particular, we have considered quadratic, Gaussian and exponential status distributions represented by red, purple and blue hues respectively. In each of these plots, the status entropy of each community in the network is plotted against $h$ and represented by $\circ$. The thick solid line is the mean status entropy at a given $h$. The dashed lines (hues of gray) represent the surrogate cases performed at different organizational scales. The slopes ($\beta$) corresponding to all these lines are mentioned in the legend. (d-f) Variation of the mean status of a community with its order $h$ where the colorbar represents depth $d$.}
    \label{fig_s2_entropy}
\end{figure}

Next, we examine the variation of $\langle E \rangle_h$, the average status entropy of communities at fixed $h$ with the scale $h$ for networks obtained from different status distributions. Clearly, at each organizational scale, the entropy is minimized by the actual structure of the network when compared to a surrogate network with randomized status distributions at that scale (see Fig. \ref{fig_s2_entropy}(a-c)). This observation is true irrespective of the underlying status distribution. Therefore, it is evident that the universality in the topology of hierarchical communities across systems arises due to the general self-organizing principle we have described in this work.

However, the distribution of mean status of communities ($\langle S\rangle_{C_j}$) with the organizational scale $h$ is remarkably different for the networks obtained from distinct status distributions. Interestingly, distribution of $\langle S\rangle_{C_j}$ with $h$ appears similar to the original status distribution of nodes. For example, see Fig. \ref{fig_s2_entropy}(d) obtained from a network for a quadratic status distribution; here, we find that larger communities at $h=7$ split into smaller communities with a broad range of $\langle S\rangle_{C_j}$ at $h=6$, as compared to that in the case of Fig. \ref{fig_s2_entropy}(e,f). In other words, the distribution of $\langle S\rangle_{C_j}$ is broad for high values of $h$ and resembles the concave and broad quadratic distribution of statuses. On the other hand, in Fig. \ref{fig_s2_entropy}(f), the distribution of $\langle S\rangle_{C_j}$ remains narrow at high values of $h$ and broadens only at lower values of $h$ similar to the underlying exponential distribution of statuses. 

We therefore infer that the composition of communities (mean status of nodes) varies remarkably for distinct status distributions. Also, the size and number of communities formed at different organizational scales can be different if the underlying status distributions of nodes are different (as observed for real-world systems in Fig. \ref{fig_s1_adj}). Yet, we find striking universality in the topology of these networks. In summary, due to non-uniformity in the status distribution, dissimilar nodes separate into different groups. Dissimilarity among nodes can be large or small resulting in communities embedded within communities at multiple scales. Irrespective of the underlying status distribution, it is the local rule of link-formation that determines the organizing principles. The self-organizing principle thus minimizes the dissimilarity of any node relative to the nodes in the same community at every scale.

\section{Effect of link density on the hierarchical community structure of the network}\label{suppsec_vary_m}

Link density plays an important role in determining the features of hierarchical communities in a network. The input parameter $m$ in our model determines the number of connections that an incoming node makes with the existing nodes in the network. As $m$ increases, the link density of the resultant network increases. The scaling exponents obtained from networks for different values of $m$ are presented in Table \ref{table_supp_vary_m}.

\begin{table}[]
\centering
\begin{tabular}{|c|c|rrrrr|}
\hline
\multirow{2}{*}{Coefficient} & \multirow{2}{*}{$y$ versus $x$}                                & \multicolumn{5}{c|}{Status model networks - varying $m$}                                                                                                                             \\ \cline{3-7} 
                             &                                                                & \multicolumn{1}{c|}{$m=2$}  & \multicolumn{1}{c|}{$m=4$}  & \multicolumn{1}{c|}{$m=8$}  & \multicolumn{1}{c|}{$m=10$} & \multicolumn{1}{c|}{Universal law} \\ \hline
$\gamma_b$                   & $\log_{10}(b_h)$ vs $h$                                        & \multicolumn{1}{r|}{\magb}  & \multicolumn{1}{r|}{\mbgb}  & \multicolumn{1}{r|}{\mdgb}  & \multicolumn{1}{r|}{\megb}  & \mgb                            \\ \hline
$\gamma_\chi$                & $\log_{10}(\chi_h)$ vs $h$                                     & \multicolumn{1}{r|}{\magx}  & \multicolumn{1}{r|}{\mbgx}  & \multicolumn{1}{r|}{\mdgx}  & \multicolumn{1}{r|}{\megx}  & \mgx                            \\ \hline
$\gamma_\eta$                & $\log_{10}(\langle \eta \rangle _h)$ vs $h$                    & \multicolumn{1}{r|}{\mage}  & \multicolumn{1}{r|}{\mbge}  & \multicolumn{1}{r|}{\mdge}  & \multicolumn{1}{r|}{\mege}  & \mge                            \\ \hline
$\gamma_n$                   & $\log_{10}(\langle n \rangle _h)$ vs $h$                       & \multicolumn{1}{r|}{\magn}  & \multicolumn{1}{r|}{\mbgn}  & \multicolumn{1}{r|}{\mdgn}  & \multicolumn{1}{r|}{\megn}  & \mgn                            \\ \hline
$\gamma_{h}$                 & $\log_{10}(\langle h \rangle _d)$ vs $d$                       & \multicolumn{1}{r|}{\magh}  & \multicolumn{1}{r|}{\mbgh}  & \multicolumn{1}{r|}{\mdgh}  & \multicolumn{1}{r|}{\megh}  & \mgh                            \\ \hline
$\gamma_{\chi_d}$            & $\log_{10}(\chi _d)$ vs $\langle h \rangle _d$                 & \multicolumn{1}{r|}{\magxd} & \multicolumn{1}{r|}{\mbgxd} & \multicolumn{1}{r|}{\mdgxd} & \multicolumn{1}{r|}{\megxd} & \mgxd                           \\ \hline
$\gamma_{\eta_d}$            & $\log_{10}(\langle \eta \rangle _d)$ vs $\langle h \rangle _d$ & \multicolumn{1}{r|}{\maged} & \multicolumn{1}{r|}{\mbged} & \multicolumn{1}{r|}{\mdged} & \multicolumn{1}{r|}{\meged} & \mged                           \\ \hline
$\gamma_{n_d}$               & $\log_{10}(\langle n \rangle _d)$ vs $\langle h \rangle _d$    & \multicolumn{1}{r|}{\magnd} & \multicolumn{1}{r|}{\mbgnd} & \multicolumn{1}{r|}{\mdgnd} & \multicolumn{1}{r|}{\megnd} & \mgnd                           \\ \hline
\end{tabular}%
\caption{Horton scaling exponents with goodness of fit (R-square values in brackets) for networks with different link densities. Networks with distinct link densities are obtained by varying the parameter $m$ in our model while keeping the other parameters constant ($N = 1000$, $N_0 = 100$ and $p(S)$ is a Gaussian with zero mean and unit variance). Average of each exponent across networks with different $m$ is reported in the last column along with the standard error with $90\%$ confidence.}\label{table_supp_vary_m}
\end{table}

\begin{figure}
    \centering
    \includegraphics[width=\textwidth]{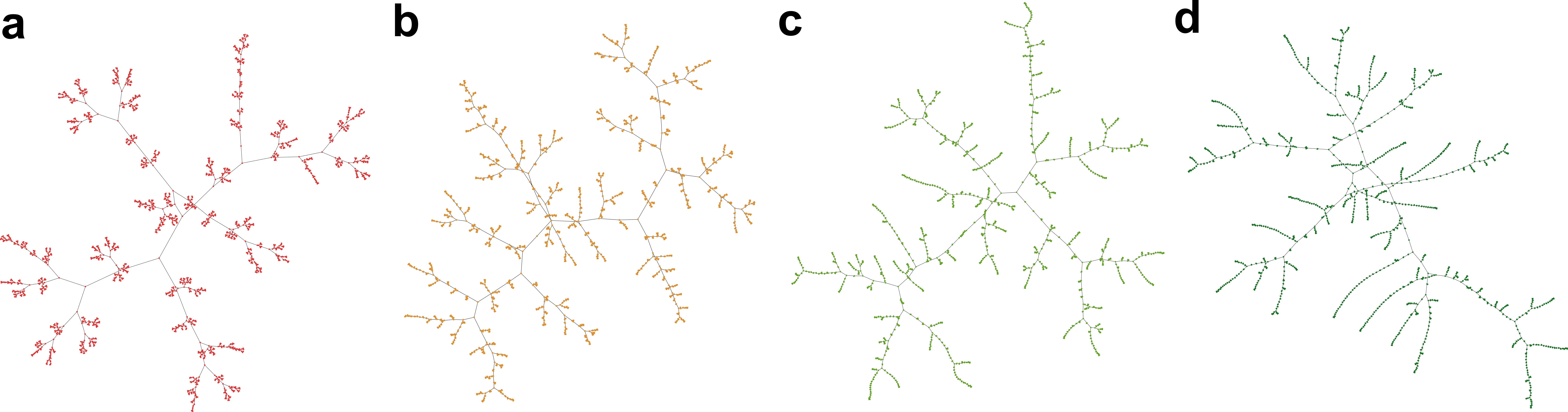}
    \caption{\textbf{Binary tree representation of the hierarchical organization of communities of different networks simulated using our model, in increasing order of link density.} In particular, we consider (a) $m=2$, (b) $m=4$, (c) $m=8$, (d) $m=10$.}
    \label{fig_s3_bt}
\end{figure}

We find that for different values of $m$, the tree representation of the resulting network exhibits a self-similar branching structure. We also notice that networks with higher link density display a branching structure with relatively longer branches at smaller organizational scales (for example, compare the tree representations for networks when $m=4$ and $m=10$ in Fig. \ref{fig_s3_bt}(b,d), respectively). As we increase the value of $m$, each node is forced to make connections with more nodes that can be less similar in status. When a large value of $m$ is chosen, connections between nodes in different groups comprising nodes with similar statuses increase. As a result, connections between communities increase and the community structure becomes less distinguishable, starting at the smaller organizational scales. 

\section{Effect of network size and growth}\label{suppsec_growth}

Here, we examine the effect of network size and growth on the hierarchical community structure of a network. We construct networks of different sizes ($N$) while preserving the link density and for the same underlying status distribution. The various scaling exponents derived from the tree representation of these networks is listed in Table \ref{table_supp_vary_N}. The changes in the value of these exponents are small for large changes in $N$. That is, we obtain networks with self-similar hierarchical communities of distinct sizes.

\begin{table}[h]
\centering
\begin{tabular}{|c|c|rrrr|}
\hline
\multirow{2}{*}{Coefficient} & \multirow{2}{*}{$y$ versus $x$}                                & \multicolumn{4}{c|}{Status model networks - varying parameter $N$}                                                                   \\ \cline{3-6} 
                             &                                                                & \multicolumn{1}{c|}{$N = 1000$} & \multicolumn{1}{c|}{$N = 2000$} & \multicolumn{1}{c|}{$N = 3000$} & \multicolumn{1}{c|}{Universal law} \\ \hline
$\gamma_b$                   & $\log_{10}(b_h)$ vs $h$                                        & \multicolumn{1}{r|}{\nagb}      & \multicolumn{1}{r|}{\nbgb}      & \multicolumn{1}{r|}{\ncgb}      & \ngb                            \\ \hline
$\gamma_\chi$                & $\log_{10}(\chi_h)$ vs $h$                                     & \multicolumn{1}{r|}{\nagx}      & \multicolumn{1}{r|}{\nbgx}      & \multicolumn{1}{r|}{\ncgx}      & \ngx                            \\ \hline
$\gamma_\eta$                & $\log_{10}(\langle \eta \rangle _h)$ vs $h$                    & \multicolumn{1}{r|}{\nage}      & \multicolumn{1}{r|}{\nbge}      & \multicolumn{1}{r|}{\ncge}      & \nge                            \\ \hline
$\gamma_n$                   & $\log_{10}(\langle n \rangle _h)$ vs $h$                       & \multicolumn{1}{r|}{\nagn}      & \multicolumn{1}{r|}{\nbgn}      & \multicolumn{1}{r|}{\ncgn}      & \ngn                            \\ \hline
$\gamma_{h}$                 & $\log_{10}(\langle h \rangle _d)$ vs $d$                       & \multicolumn{1}{r|}{\nagh}      & \multicolumn{1}{r|}{\nbgh}      & \multicolumn{1}{r|}{\ncgh}      & \ngh                            \\ \hline
$\gamma_{\chi_d}$            & $\log_{10}(\chi _d)$ vs $\langle h \rangle _d$                 & \multicolumn{1}{r|}{\nagxd}     & \multicolumn{1}{r|}{\nbgxd}     & \multicolumn{1}{r|}{\ncgxd}     & \ngxd                           \\ \hline
$\gamma_{\eta_d}$            & $\log_{10}(\langle \eta \rangle _d)$ vs $\langle h \rangle _d$ & \multicolumn{1}{r|}{\naged}     & \multicolumn{1}{r|}{\nbged}     & \multicolumn{1}{r|}{\ncged}     & \nged                           \\ \hline
$\gamma_{n_d}$               & $\log_{10}(\langle n \rangle _d)$ vs $\langle h \rangle _d$    & \multicolumn{1}{r|}{\nagnd}     & \multicolumn{1}{r|}{\nbgnd}     & \multicolumn{1}{r|}{\ncgnd}     & \ngnd                           \\ \hline
\end{tabular}%
\caption{Horton scaling exponents with goodness of fit (R-square values in brackets) for different model networks with varying size ($N$). We consider networks with $N = 1000$, $N = 2000$ and $N = 3000$ with $m = 4$, $m = 7$ and $m = 11$ respectively. This ensures that the link density across these networks remains roughly the same. Other parameters are $N_0 = 100$ and $p(S)$ is a Gaussian ($\mu = 0$, $\sigma=1$). The mean exponents for all these cases is also mentioned in the last column along with the standard error ($90\%$ confidence).}
\label{table_supp_vary_N}
\end{table}

Next, we investigate the effect of growth in determining the structure of the network. We present two models $M_1$ and $M_2$ where a network is formed using the same principle as our model in the main manuscript; i.e., nodes that are more similar to each other are more probable to connect. However, in these models we do not add nodes at every time step, the networks are non-growing.

\subsubsection*{Non-growing network model $M_1$}

In model $M_1$, we predefine the total number of nodes $N$ and the status distribution $p(S)$. We consider all possible pairs of nodes $\{i,j\}$ and form a connection based on a probability $\pi_{ij}$. The probability $\pi_{ij}$ that any two nodes $i$ and $j$ are connected is given by Eq. \ref{eq_probab}. 
\begin{equation}\label{eq_probab}
    \pi _{ij} = \frac{d_{ij}}{\sum_k d_{ik}}    
\end{equation}
Here, $d_{ij}$ is the reciprocal of the difference in the statuses of two nodes $i$ and $j$, i.e., $d_{ij}={1}/|{S_i - S_j}|$. Note that, the link density of a network constructed using this model is not predefined. 

\begin{figure}
    \centering
    \includegraphics[width=0.65\textwidth]{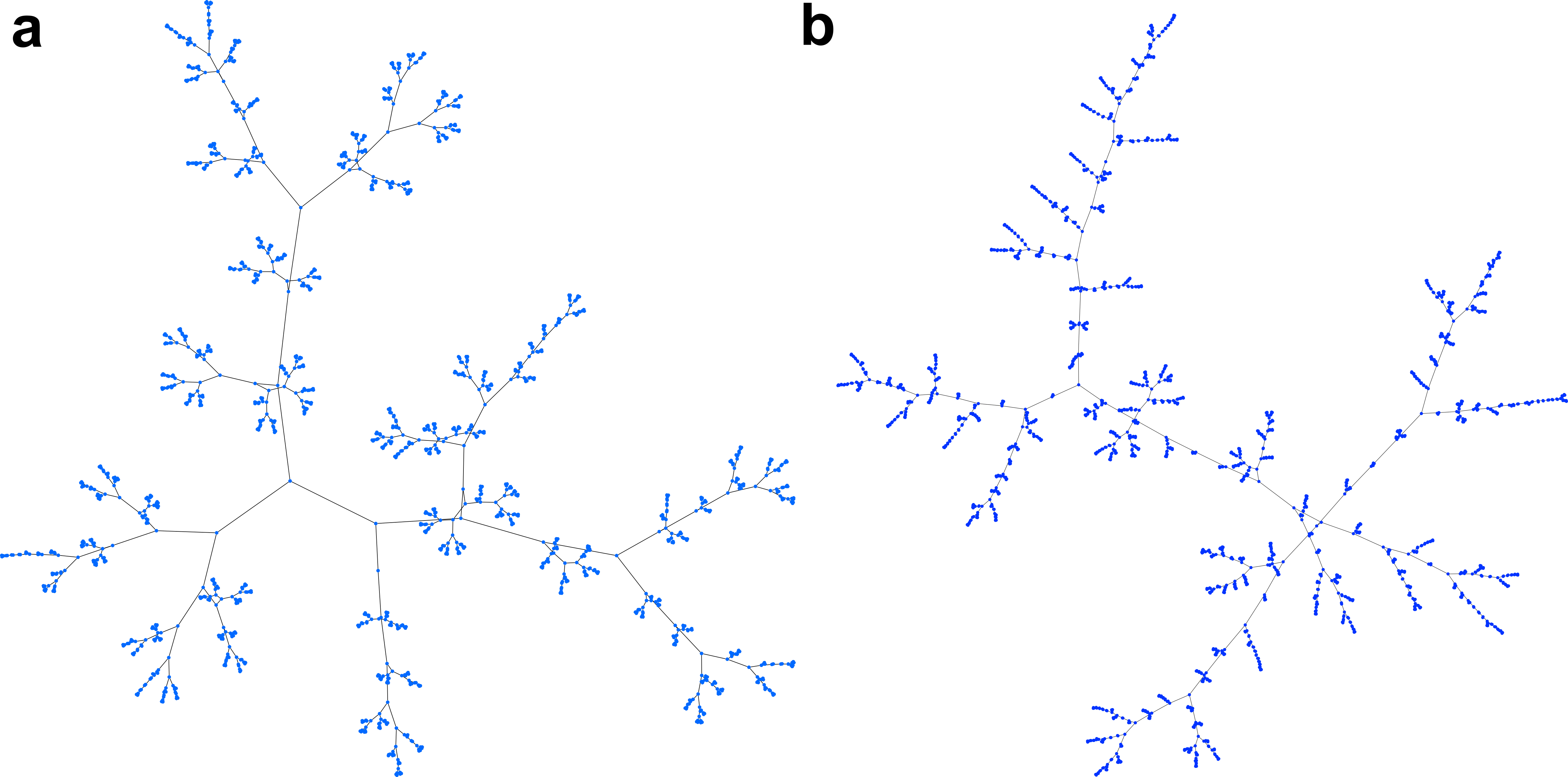}
    \caption{\textbf{Self-similar tree representations of networks obtained from non-growing network models.} Binary trees obtained from networks simulated using non-growing network models: (a) model $M_1$, (b) model $M_2$.}
    \label{fig_supp_M1M2_bt}
\end{figure}

For $N=1000$ and $p(S)$ as a Gaussian probability distribution with zero mean and unit variance, we find that the resulting network comprises hierarchical communities. Further, the branching in the tree representation appears to be self-similar (see Fig. \ref{fig_supp_M1M2_bt}(a)). The scaling exponents are reported in Table \ref{table_supp_growth_models}. Evidently, the hierarchical community structure of the network obtained from model M1 is self-similar.

\subsubsection*{Non-growing network model $M_2$}

In model $M_2$, we construct a network of a total number of $N$ nodes and an underlying status distribution $p(S)$. The links between nodes are established in the following manner. Each node must make at least $m$ connections with other nodes that are selected randomly. The probability of forming a connection is the same as Eq. \ref{eq_probab}. In this model, we fix the minimum number of connections that each node has. 

Interestingly, for a status distribution that is a Gaussian, $N=1000$, $m=4$ the resulting network comprises hierarchical communities and the branching structure of the tree representation appears self-similar (in Fig. \ref{fig_supp_M1M2_bt}(b)). The scaling exponents are reported in Table \ref{table_supp_growth_models}. Evidently, the scaling exponents are very different for networks formed using the growth-based model in the main manuscript and the model $M_2$. These differences arise owing to the differences in the sample space of other nodes in the network available to a node for forming links. 
\begin{table}[]
\centering
\begin{tabular}{|c|c|rr|}
\hline
\multirow{2}{*}{Coefficient} & \multirow{2}{*}{$y$ versus $x$}                                & \multicolumn{2}{c|}{\begin{tabular}[c]{@{}c@{}}Non-growing network \\models\end{tabular}} \\ \cline{3-4} 
                             &                                                                & \multicolumn{1}{c|}{$M_1$}                    & \multicolumn{1}{c|}{$M_2$}                    \\ \hline
$\gamma_b$                   & $\log_{10}(b_h)$ vs $h$                                        & \multicolumn{1}{r|}{\npsgb}                        & \nmsgb                                             \\ \hline
$\gamma_\chi$                & $\log_{10}(\chi_h)$ vs $h$                                     & \multicolumn{1}{r|}{\npsgx}                        & \nmsgx                                             \\ \hline
$\gamma_\eta$                & $\log_{10}(\langle \eta \rangle _h)$ vs $h$                    & \multicolumn{1}{r|}{\npsge}                        & \nmsge                                             \\ \hline
$\gamma_n$                   & $\log_{10}(\langle n \rangle _h)$ vs $h$                       & \multicolumn{1}{r|}{\npsgn}                        & \nmsgn                                             \\ \hline
$\gamma_{h}$                 & $\log_{10}(\langle h \rangle _d)$ vs $d$                       & \multicolumn{1}{r|}{\npsgh}                        & \nmsgh                                             \\ \hline
$\gamma_{\chi_d}$            & $\log_{10}(\chi _d)$ vs $\langle h \rangle _d$                 & \multicolumn{1}{r|}{\npsgxd}                       & \nmsgxd                                            \\ \hline
$\gamma_{\eta_d}$            & $\log_{10}(\langle \eta \rangle _d)$ vs $\langle h \rangle _d$ & \multicolumn{1}{r|}{\npsged}                       & \nmsged                                            \\ \hline
$\gamma_{n_d}$               & $\log_{10}(\langle n \rangle _d)$ vs $\langle h \rangle _d$    & \multicolumn{1}{r|}{\npsgnd}                       & \nmsgnd                                            \\ \hline
\end{tabular}
\caption{Horton scaling exponents with goodness of fit (R-square values in brackets) for networks obtained from two different non-growing versions of status model. We consider two networks, each with $N=1000$ and Gaussian status distribution (zero mean and unit variance), one obtained using $M_1$ and the other through $M_2$, with $m=4$.}
\label{table_supp_growth_models}
\end{table}

In a growing model, a newly entrant node can make $m$ connections with the nodes already added to the network; whereas, in model $M_2$, a node makes $m$ connections with the entire set of $N$ nodes in the network. As a result, the quantitative features of the topology of hierarchical communities differ between the two approaches. Yet, we observe that growth of a network is not a key ingredient for the formation of hierarchical communities.

 \end{document}